\documentclass[12pt]{article}
\usepackage{psfrag}
\usepackage{cite}
\usepackage{a4wide}
\usepackage{amsmath,amsthm}
\usepackage{graphicx}
\usepackage{amssymb}
\usepackage{amsopn}
\usepackage{color}
\allowdisplaybreaks

\makeatletter \@addtoreset{equation}{section} \makeatother

\begin{document}
{\renewcommand{\thefootnote}{\fnsymbol{footnote}}
\begin{center}
{\LARGE  Quasi-Local Energy in Loop Quantum Gravity}\\
\vspace{1.5em} Jinsong Yang\footnote{e-mail address: {\tt
yangksong@gmail.com}} and  Yongge Ma\footnote{
e-mail address: {\tt mayg@bnu.edu.cn}}
\\
\vspace{0.5em} Department of Physics, Beijing Normal University,
Beijing 100875, China\vspace{1.5em}
\end{center}
}

\setcounter{footnote}{0}

\newcommand{\Lam}{\Lambda}
\newcommand{\vt}{\vartheta}
\newcommand{\vp}{\varphi}
\newcommand{\abs}[1]{\lvert#1\rvert}

\newtheorem{theo}{Theorem}[section]
\newtheorem{defi}{Definition}[section]

\newcommand{\case}[2]{{\textstyle \frac{#1}{#2}}}
\newcommand{\lP}{\ell_{\mathrm P}}

\newcommand{\md}{{\mathrm{d}}}
\newcommand{\Kern}{\mathop{\mathrm{ker}}}
\newcommand{\tr}{\mathop{\mathrm{tr}}}
\newcommand{\sgn}{\mathop{\mathrm{sgn}}\nolimits}

\newcommand*{\R}{{\mathbb R}}
\newcommand*{\N}{{\mathbb N}}
\newcommand*{\Z}{{\mathbb Z}}
\newcommand*{\Q}{{\mathbb Q}}
\newcommand*{\C}{{\mathbb C}}

\date{28 August 2008}

\begin{abstract}
Although there is no known meaningful notion of the energy density
of the gravitational field in general relativity, a few notions of
quasi-local energy of gravity associated to extended but finite
domains have been proposed. In this paper, the notions of
quasi-local energy are studied in the framework of loop quantum
gravity, in order to see whether these notions can be carried out
at quantum level. Two basic quasi-local geometric quantities are
quantized, which lead to well-defined operators in the kinematical
Hilbert space of loop quantum gravity. We then use them as basic
building blocks to construct different versions of quasi-local
energy operators. The operators corresponding to Brown-York
energy, Liu-Yau energy, Hawking energy, and Geroch energy are
obtained respectively. The virtue of the Geroch energy operator is
beneficial for us to derive a rather general entropy-area relation
and thus a holographic principle from loop quantum gravity.

\end{abstract}

{PACS number(s): 04.60.Pp, 04.20.Cv}

\section{Introduction}

It is well known that there are inherent difficulties in defining
energy in general relativity (GR), essentially owing to its
non-localizability. By now there is no known meaningful notion of
the energy density of gravitational field in GR. Globally, for
spacetimes which are asymptotically flat, there are well-defined
notions for the total energy, given by the Bondi and ADM
expressions integrated over spheres at null infinity and spatial
infinity. These global notions are directly related to quantities
that can be measured physically by distant observers. However,
finding an appropriate notion of energy-momentum would be
important from the point of view of applications as will. For
example, the correct, ultimate formulation of black hole
thermodynamics should probably be based on quasi-local defined
internal energy, entropy, angular momentum etc. So far,
considerable efforts have been put in to formulate a satisfactory
definition of quasi-local energy (QLE) (see \cite{QLEreview} for a
review). In this paper, a few expressions of quasi-local energy
are quantized in the framework of loop quantum gravity (LQG) (see
\cite{ThomasRev,RovRev,ALRev,HanRev} for reviews). Our purpose is
in two folds. Firstly, we want to check whether the quasi-local
notions of gravitational energy can be carried out at quantum
level. Secondly, we wish to use these notions of quantum
gravitational energy to study the relation between quantum gravity
and gravitational thermodynamics. A similar effect was made in
\cite{MarjorOp} to quantize the Hamiltonian surface term of a
bounded spatial region. The candidates of QLE that we are
considering include the Brown-York energy \cite{Brown-York},
Liu-Yau energy \cite{Liu-Yau}, Hawking energy \cite{Hawking}, and
Geroch energy \cite{GE}. All of these energy expressions are
constituted by two basic quasi-local quantities representing the
extrinsic curvatures of an two-sphere in a spatial slice and a
timelike slice respectively. Thus the key task is to quantize
these two basic building blocks.

In Section 2, the kinematical framework of LQG is briefly
introduced. We then construct in section 3 two basic operators in
the kinematical Hilbert space of LQG, which corresponding to the
two basic building blocks of quasi-local energies. In section 4,
the two basic operators are employed to construct different
versions of QLE operators. In section 5, the Geroch energy
operator is used to derive a rather general entropy-area relation
and thus a holographic principle from LQG.

\section{Elements of LQG}

The Hamiltonian formalism of GR is formulated on a 4-dimensional
manifold $M=\mathbb{R}\times\Sigma$, where $\Sigma$ represents a
3-dimensional manifold with arbitrary topology. Introducing
Ashtekar-Barbero variables \cite{AshVar,Barbero}, GR can be casted
into an $SU(2)$ connection dynamical theory. The phase space
consists of canonical pairs $(A^i_a, E^a_i)$ of fields on $\Sigma$,
where $A^i_a$ is a connection 1-form which takes values in the Lie
algebra $su(2)$, and $E^a_i$ is a vector density of weight 1 which
takes value in the dual of $su(2)$. Here $a,b,c...$ are abstract
spatial indices and $i,j,k...=1,2,3$ are internal $su(2)$-indices.
The density-weighted triad $E^a_i$ is related to the co-triad
$e_a^i$ by the relation
$E^a_i=\frac{1}{2}\,\epsilon^{abc}\epsilon_{ijk}e^j_be^k_c\sgn(\det(e^i_a))$,
where $\epsilon^{abc}$ is the naturally defined levi-civita density
and $\sgn(\det(e^i_a))$ denotes the sign of $\det(e^i_a)$. The
3-metric on $\Sigma$ is related to the co-triad by
$q_{ab}=e^i_ae^j_b\delta_{ij}$. The only non-trivial Poisson bracket
is given by
\begin{align}
\{A^i_a(x),E^b_j(y)\}=\kappa\beta\delta^b_a\delta^i_j\delta^3(x,y),
\end{align}
where $\kappa=8\pi G$ ($G$ denotes Newton's constant) and $\beta$
is the Barbero-Immirzi parameter. There are three first-class
constraints in this Hamiltonian formalism of gravity:
\begin{align}
G_i&={\cal D}_aE^a_i=\partial_aE^a_i+\epsilon_{ijk}A^j_aE^a_k,\nonumber\\
V_a&=F_{ab}^iE^b_i,\nonumber\\
H&=\frac{E^{aj}E^{bk}}{2\kappa\sqrt{|\det(q)|}}\big[\epsilon_{ijk}F_{ab}^i-(1+\beta^2)2K^j_{[a}K^k_{b]})\big],
\end{align}
where ${\cal D}_a$ denotes the covariant derivative defined by the
connection $A^i_a$,
$F^i_{ab}:=\partial_aA^i_b-\partial_bA^i_a+{\epsilon^i}_{jk}A^j_aA^k_b$
is the curvature of $A^i_a$, and $K_a^i$ is the extrinsic curvature
of $\Sigma$.

One element of LQG is the notion of graphs embedded in $\Sigma$.
By $\gamma$ we denote a closed, piecewise analytic graph. The set
of edges of $\gamma$ is denoted by $E(\gamma)$ and the set of
vertices of $\gamma$ by $V(\gamma)$. For an oriented edge $e$ of
$\gamma$, its beginning point is denoted by $b(e)$ and its final
point by $f(e)$. To construct quantum kinematics, one has to
extend the configuration space ${\cal A}$ of smooth connections to
the space $\bar{\cal A}$ of distributional connections. Through
projective techniques, $\bar{\cal A}$ is equipped with a natural,
faithful, `induced' measure $\mu_o$, called
Ashtekar-Isham-Lewandowski measure \cite{AI,AL}. In certain sense,
this measure is the unique diffeomorphism-invariant measure on
$\bar{\cal A}$ \cite{Uniq}. The kinematical Hilbert space then
reads ${\cal H}_{\mathrm{kin}}=L^2(\bar{\cal A},d\mu_o)$. The
so-called spin-network basis $T_{\gamma,j,m,n}$ provide an
orthonormal basis for ${\cal H}_{\mathrm{kin}}$ \cite{ThomasRev}.

One of successes of LQG is the rigorous construction of spatial
geometrical operators, such as the area, the volume and the length
operators in ${\cal H}_{\mathrm{kin}}$ \cite{Area,Volum,Length}.
Moreover it turns out that these geometrical operators have a
discrete spectrum. The same conclusion are also tenable in the
internal gauge invariant Hilbert space ${\cal
H}_o=L^2(\overline{\cal{A/G}},d\mu_o)$.

\section{Two basic operators for QLE}

Most of the quasi-local energy expressions appeared so far involve
two quasi-local quantities defined by the integrals of two extrinsic
scalar curvatures of some spatial 2-surface \cite{QLEreview}. In
this section, we will construct two well-defined basic operators
corresponding to the two quasi-local quantities in the kinematic
Hilbert space ${\cal H}_{\mathrm{kin}}$ of LQG, which will be used
as basic building blocks to construct different versions of
quasi-local energy operators in the next section. To this aim, we
have to first re-express the two quasi-local quantities in terms of
real connection variable or its conjugate. Then we regulate the
classical expressions in order to get quantities with quantum
analogues. It turns out that in the regularization procedure, as the
regularization of the Hamiltonian constraint, we need to triangulate
the 3-d spatial manifold $\Sigma$ in adaption to a graph, which
comes from the cylindrical function in ${\cal H}_{\mathrm{kin}}$
that is going to be acted by the constructed operators.

\subsection{QLE-like operator}\label{section1}

Let $S$ be a 2-d surface with two-sphere topology in the 3-d
spatial manifold $\Sigma$ and $\sigma_{ab}$ be the induced metric
on $S$ of metric $q_{ab}$ on $\Sigma$. For simplicity, we choose
adapted coordinates $\{x^1,x^2,x^3\}$ in $\Sigma$ such that $S$ is
given by $x^3=0$, and $x^1,x^2$ parameterize $S$. The QLE-like
observable is defined as
\begin{align}
\label{D1} E_{Q,k}(S):=-\frac{1}{\kappa}\int_S
d^2x\sqrt{\det(\sigma)}\;k,
\end{align}
where $k$ is the extrinsic scalar curvature of $k_{ab}$ of $S$
corresponding to the unit normal $n^a$ in $\Sigma$. In order to
quantize the expression, we need first to express it in terms of
the real Ashtekar variables. The extrinsic curvature tensor
$k_{ab}$ of $S$ corresponding to $n^a$ reads
\begin{align}
k_{ab}={\sigma_a}^cD_cn_b,
\end{align}
where $D_a$ is the derivative operator on $\Sigma$ compatible with
$q_{ab}$, i.e., $D_aq_{bc}=0$. So the extrinsic scalar curvature
$k$ of $S$ is
\begin{align}
\label{classk} k=&D_an^a.
\end{align}
From (\ref{classk}), we have the following identity in the adapted
coordinates.
\begin{align}
\label{kAE}
k&=D_a\left(\frac{E^a_iE^b_i}{\det(q)}n_b\right)=\frac{1}{\sqrt{\det(q)}}\;\partial_a
\left(\frac{E^a_iE^b_i}{\sqrt{\det(q)}}n_b\right)
=\frac{1}{\sqrt{\det(q)}}\;\partial_a\left(\frac{E^3_iE^a_i}{\sqrt{\det(\sigma)}}\right).
\end{align}
To regularize $E_{Q,k}(S)$, let $\epsilon$ be a small number and
$\chi^3_{\epsilon}(x,y)$ be the (smoothed out) characteristic
function such that
$\lim_{\epsilon\rightarrow0}\chi^3_\epsilon(x,y)/\epsilon^3=\delta^3(x,y)$.
The volume of the cube as measured by $q_{ab}$ is given by
$V(x,\epsilon):=\int d^3y\chi^3_\epsilon(x,y)\sqrt{\det(q)}(y)$ such
that $\lim_{\epsilon\rightarrow
0}\frac{V(x,\epsilon)}{\epsilon^3}=\sqrt{\det(q)}(x)$. We choose
again adapted coordinates $\{y^1,y^2,y^3\}$ in $\Sigma$ such that
$S$ is given by $y^3=0$ and each 2-d surface $S_{y^3}$ of the family
given by $y^3=constant$ is parameterized by $y^1,y^2$. We denote
also the induced metric of $q_{ab}$ in $S_{y^3}$ by $\sigma_{ab}$.
We then have the following identity by Eqs. (\ref{D1}) and
(\ref{kAE})
\begin{align}
\label{Ek1}
E_{Q,k}(S)
 &=-\frac{1}{\kappa}\int_S d^2x\frac{\sqrt{\det(\sigma)}}{\sqrt{\det(q)}}\;\partial_a
 \left(\frac{E^3_iE^a_i}{\sqrt{\det(\sigma)}}\right)\nonumber\\
 &=-\lim_{\epsilon\rightarrow 0}\frac{1}{\kappa}\int_S d^2x\sqrt{\det(\sigma)}(x)\int
    dy^3\int_{S_{y^3}}dy^1dy^2\,\frac{\chi^3_\epsilon(x,y)}{\epsilon^3}\;\partial_a
    \left(\frac{E^3_iE^a_i}{\sqrt{\det(\sigma)}}(y)\right)\nonumber\\
 &\quad\quad\quad\quad\times\int_\Sigma
   d^3u\frac{\chi^3_\epsilon(x,u)}{\epsilon^3}\frac{1}{\sqrt{\det(q)}(u)}\nonumber\\
 &=-\lim_{\epsilon\rightarrow 0}\frac{1}{\kappa}\int_S d^2x\sqrt{\det(\sigma)}(x)\int
    dy^3\int_{S_{y^3}}dy^1dy^2\,\chi^3_\epsilon(x,y)\;\partial_a
    \left(\frac{E^3_iE^a_i}{\sqrt{\det(\sigma)}}(y)\right)\nonumber\\
 &\quad\quad\times\int_\Sigma
    d^3u\chi^3_\epsilon(x,u)\frac{\left[\det(e^i_b)\right](u)}{[\sqrt{V(u,\epsilon)}\;]^3}
    \int_\Sigma d^3w\chi^3_\epsilon(x,w)
    \frac{\left[\det(e^i_b)\right](w)}{[\sqrt{V(w,\epsilon)}\;]^3}\nonumber\\
 &=\lim_{\epsilon\rightarrow 0}\frac{1}{\kappa}\int_S
d^2x\sqrt{\det(\sigma)}(x)\int
    dy^3\int_{S_{y^3}}dy^1dy^2\,[\partial_a\chi^3_\epsilon(x,y)]\;
    \left(\frac{E^3_iE^a_i}{\sqrt{\det(\sigma)}}(y)\right)\nonumber\\
 &\quad\quad\times\int_\Sigma
    d^3u\chi^3_\epsilon(x,u)\frac{\left[\det(e^i_b)\right](u)}{[\sqrt{V(u,\epsilon)}\;]^3}
    \int_\Sigma d^3w\chi^3_\epsilon(x,w)
    \frac{\left[\det(e^i_b)\right](w)}{[\sqrt{V(w,\epsilon)}\;]^3}\,,
\end{align}
where we have inserted the identity
$1=\left[\det(e_a^i)\right]^2/[\sqrt{\det(q)}\;]^2$ in the third
step, and performed an integration by parts in the last step . Let
$\chi^2_{\epsilon'}(y,z)$ is the 2-d characteristic function of a
coordinate box with center $y$ and coordinate area ${\epsilon'}^2$
and $Ar(z,\epsilon'):=\int_{S_{y^3}}
d^2w\chi^2_{\epsilon'}(z,w)\sqrt{\det(\sigma)}(w)$, satisfying
$\lim_{\epsilon'\rightarrow
0}\frac{Ar(z,\epsilon')}{{\epsilon'}^2}=\sqrt{\det(\sigma)}(z)$, is
the area of the box as measured by $\sigma_{ab}$. Then we have
\begin{align}
\frac{E^3_i}{\sqrt{\det(\sigma)}}(y^1,y^2,y^3)&=\lim_{\epsilon'\rightarrow0}\int_{S_{y^3}}d^2z
\frac{\chi^2_{\epsilon'}(y^1,y^2;z^1,z^2)}
{\epsilon'^2}\frac{E^3_i}{\sqrt{\det(\sigma)}}(z^1,z^2,y^3)\nonumber\\
&=\lim_{\epsilon'\rightarrow0}\int_{S_{y^3}}d^2z\frac{\chi^2_{\epsilon'}(y^1,y^2;z^1,z^2)E^3_i(z^1,z^2,y^3)}
{Ar(z^1,z^2,y^3;\epsilon')}\,.\nonumber
\end{align}
Thus we can rewrite Eq. (\ref{Ek1}) as
\begin{align}
E_{Q,k}(S)
 &=\lim_{\epsilon,\epsilon'\rightarrow 0}\frac{1}{\kappa}\int_S
d^2x\sqrt{\det(\sigma)}(x)\nonumber\\
&\quad\quad\times\int_\Sigma
    d^3y\left[\int_{S_{y^3}}d^2z\frac{\chi^2_{\epsilon'}(y^1,y^2;z^1,z^2)E^3_i(z^1,z^2,y^3)}
   {Ar(z^1,z^2,y^3;\epsilon')}\right]\left[\partial_a\chi^3_\epsilon(x,y)\right]E^a_i(y)\nonumber\\
 &\quad\quad\times\int_\Sigma
    d^3u\chi^3_\epsilon(x,u)\frac{\left[\det(e^i_b)\right](u)}{[\sqrt{V(u,\epsilon)}\;]^3}
    \int_\Sigma d^3w\chi^3_\epsilon(x,w)
    \frac{\left[\det(e^i_b)\right](w)}{[\sqrt{V(w,\epsilon)}\;]^3}\,.\label{E_Q}
\end{align}
Recall the following classical identities
\begin{align}
&\int_\Sigma
d^3u\left[\det(e_b^i)\right](u)=\frac{1}{3!}\int_\Sigma\epsilon_{ijk}e^i\wedge
e^j\wedge e^k=\frac{-4}{3!}\int_\Sigma\mathrm{Tr}(e\wedge e\wedge
e),
\end{align}
and
\begin{align}
&e_a^i(u)=\frac{2}{\kappa}\,\{A_a^i(u),V(u,\epsilon)\},
\end{align}
where
$\epsilon_{ijk}=-\frac{1}{2}\,\mathrm{tr}(\tau_i\tau_j\tau_k)$,
$e=e^i\tau_i/2$, here $\tau_i=-i\sigma_i$ ($\sigma_i$ is the Pauli
matrix) is the generator of $su(2)$ obeying
$[\tau_i,\tau_j]=2\,\epsilon_{ijk}\tau_k$. We can rewrite Eq.
(\ref{E_Q}) as
\begin{align}
\label{regE1} E_{Q,k}(S)
 &=\frac{1}{\kappa}\left[\frac{-4}{3!}\cdot\Big(\frac{2}{\kappa}\Big)^3\right]^2\lim_{\epsilon,\epsilon'\rightarrow 0}
 \int_S d^2x\sqrt{\det(\sigma)}(x)\nonumber\\
&\quad\quad\times\int_\Sigma
    d^3y\left[\int_{S_{y^3}}d^2z\frac{\chi^2_{\epsilon'}(y^1,y^2;z^1,z^2)E^3_i(z^1,z^2,y^3)}
   {Ar(z^1,z^2,y^3;\epsilon')}\right]\left[\partial_a\chi^3_\epsilon(x,y)\right]E^a_i(y)\nonumber\\
 &\quad\quad\times\int_\Sigma
    \chi^3_\epsilon(x,u)\mathrm{Tr}\left(\frac{\{A(u),V(u,\epsilon)\}}{\sqrt{V(u,\epsilon)}}
    \wedge\frac{\{A(u),V(u,\epsilon)\}}{\sqrt{V(u,\epsilon)}}\wedge\frac{\{A(u),V(u,\epsilon)\}}
    {\sqrt{V(u,\epsilon)}}\right)\nonumber\\
 &\quad\quad\times\int_\Sigma\chi^3_\epsilon(x,w)
    \mathrm{Tr}\left(\frac{\{A(w),V(w,\epsilon)\}}{\sqrt{V(w,\epsilon)}}
    \wedge\frac{\{A(w),V(w,\epsilon)\}}{\sqrt{V(w,\epsilon)}}\wedge\frac{\{A(w),V(w,\epsilon)\}}
    {\sqrt{V(w,\epsilon)}}\right)\nonumber\\
 &=\frac{2^{14}}{9\kappa^7}\lim_{\epsilon,\epsilon'\rightarrow 0}\int_S d^2x\sqrt{\det(\sigma)}(x)\nonumber\\
&\quad\quad\times\int_\Sigma
    d^3y\left[\int_{S_{y^3}}d^2z\frac{\chi^2_{\epsilon'}(y^1,y^2;z^1,z^2)E^3_i(z^1,z^2,y^3)}
   {Ar(z^1,z^2,y^3;\epsilon')}\right]\left[\partial_a\chi^3_\epsilon(x,y)\right]E^a_i(y)\nonumber\\
 &\quad\quad\times\int_\Sigma
    \chi^3_\epsilon(x,u)\mathrm{Tr}\left(\{A(u),\sqrt{V(u,\epsilon)}\;\}
    \wedge\{A(u),\sqrt{V(u,\epsilon)}\;\}\wedge\{A(u),\sqrt{V(u,\epsilon)}\;\}\right)\nonumber\\
 &\quad\quad\times\int_\Sigma\chi^3_\epsilon(x,w)
    \mathrm{Tr}\left(\{A(w),\sqrt{V(w,\epsilon)}\}
    \wedge\{A(w),\sqrt{V(w,\epsilon)}\;\}\wedge\{A(w),\sqrt{V(w,\epsilon)}\;\}\right)\nonumber\\
 &=:\lim_{\epsilon,\epsilon'\rightarrow
 0}E^{\epsilon,\epsilon'}_{Q,k}(S),
\end{align}
where we have used
$\{\cdot,\sqrt{V(u,\epsilon)}\}=\{\cdot,V(u,\epsilon)\}/(2\sqrt{V(u,\epsilon)}\;)$.
Since a typical state $f_\gamma\in{\cal H}_{\mathrm{kin}}$ is some
cylindrical function over a graph $\gamma$ in $\Sigma$, as in the
construction of the Hamiltonian constraint operator
\cite{ThomasRev,QSDI}, we triangulate $\Sigma$ in adaption to
$\gamma$ as follows. At every vertex $v\in V(\gamma)$ we choose a
triple $(e_I, e_J, e_K)$ of edges of $\gamma$ and a tetrahedron
$\Delta_{\gamma,v,e_I,e_J,e_K}^\epsilon$ based at $v$ which is
spanned by segments $s_I, s_J, s_K$ of the triple. Each segment
$s_I$ is given by the part with the curve parameter
$t^I\in[0,\epsilon]$ of the corresponding edge $e_I(t^I)$. The
holonomy of the connection along a segment $s_I$ reads
\begin{align}
h_{s_I}(A)=\mathbb{I}_2+\epsilon\dot{s}_I^a(0)A^i_a(v)\,\tau_i/2+O(\epsilon^2),
\end{align}
and for one segment $s_I$, we have
\begin{align}
\int_{s_I}\{A(u),\sqrt{V(u,\epsilon)}\}\approx\epsilon\dot{s}^a_I(0)\{A_a(v),\sqrt{V(u,\epsilon)\,}\}
\end{align}
up to $O(\epsilon^2)$. Hence for each
$\Delta_{\gamma,v,e_I,e_J,e_K}^\epsilon$, we have
\begin{align}
&\int_{\Delta_{\gamma,v,e_I,e_J,e_K}^\epsilon}
    \mathrm{Tr}\left(\left\{A(u),\sqrt{V(u,\epsilon)}\;\right\}
    \wedge\left\{A(u),\sqrt{V(u,\epsilon)}\;\right\}\wedge\left\{A(u),\sqrt{V(u,\epsilon)}\;\right\}\right)\nonumber\\
&\approx-\frac{1}{6}\epsilon(s_Is_Js_K)\epsilon^{IJK}\mathrm{Tr}\Big(h_{s_I(\Delta)}
\left\{h_{s_I(\Delta)}^{-1},\sqrt{V(v(\Delta),\epsilon)}\;\right\}
   h_{s_J(\Delta)}\left\{h_{s_J(\Delta)}^{-1},\sqrt{V(v(\Delta),\epsilon)}\;\right\}\nonumber\\
 &\quad\quad\quad\quad\quad\quad\quad\quad\quad\quad\quad\quad\times
 h_{s_K(\Delta)}\left\{h_{s_K(\Delta)}^{-1},\sqrt{V(v(\Delta),\epsilon)}\;\right\}\Big),\nonumber
\end{align}
where $\epsilon(s_I s_J
s_K):=\mathrm{sgn}(\det(\dot{s}_I\dot{s}_J\dot{s}_K)(v))$ takes
the values $+1, -1, 0$ if the tangents of the three segments
$s_I,s_J,s_K$ at $v$ (in that sequence) form a matrix of positive,
negative or vanishing determinant. Then the integration over
$\Sigma$ can be split as follows \cite{ThomasRev,QSDI}:
\begin{align}
\label{triangle}
\int_\Sigma&=\int_{\bar{U}_\gamma^\epsilon}+\sum_{v\in
V(\gamma)}\int_{U^\epsilon_{\gamma,
v}}=\int_{\bar{U}_\gamma^\epsilon}+\sum_{v\in
V(\gamma)}\frac{1}{E(v)}\sum_{b(e_I)\cap b(e_J)\cap
b(e_K)=v}\left[\int_{U^\epsilon_{\gamma,v,e_I,e_J
,e_K}}+\int_{\bar{U}^\epsilon_{\gamma,v,e_I,e_J
,e_K}}\right]\nonumber\\
&\approx\int_{\bar{U}_\gamma^\epsilon}+\sum_{v\in
V(\gamma)}\frac{1}{E(v)}\sum_{b(e_I)\cap b(e_J)\cap
b(e_K)=v}\left[8\cdot\int_{\Delta^\epsilon_{\gamma,v,e_I,e_J,
e_K}}+\int_{\bar{U}^\epsilon_{\gamma,v,e_I,e_J,e_K}}\right].
\end{align}
Here we have first decomposed $\Sigma$ into a region
$\bar{U}^\epsilon_{\gamma}$ not containing the vertices of $\gamma$
and the regions $U^\epsilon_{\gamma, v}$ around the vertices. Then
choose a triple $(e_I, e_J, e_K)$ of edges outgoing from $v$ and
decompose $U^\epsilon_{\gamma, v}$ into the region
$U^\epsilon_{\gamma, v, e_I, e_J, e_K}$ covered by the tetrahedron
$\Delta^\epsilon_{\gamma, v, e_I, e_J, e_K}$ spanned by $e_I, e_J,
e_K$ and its 8 mirror images and the rest $\bar{U}^\epsilon_{\gamma,
v, e_I, e_J, e_K}$ not containing $v$. Note that the integral over
$U^\epsilon_{\gamma, v, e_I, e_J, e_K}$ classically converges to 8
times the integral over the original single tetrahedron
$\Delta^\epsilon_{\gamma, v, e_I, e_J, e_K}$ as we shrink the
tetrahedron to zero. We average over all such triples $(e_I, e_J,
e_K)$ and divide by the number of possible choices of triples for a
vertex $v$ with $n(v)$ edges, $E(v)=\Big({n(v)\atop 3}\Big)$. We can
now decompose the $u$ and $w$ integration over $\Sigma$ in Eq.
(\ref{regE1}) according to Eq. (\ref{triangle}). To quantize the two
integration, we replace Poisson brackets by commutators times
$1/(i\hbar)$, holonomies by multiplication and $V$ by the volume
operator $\hat{V}$, which acts on a function cylindrical over a
graph $\gamma$ as follows \cite{Volum}:
\begin{align}
\hat{V}(R)f_\gamma=(\hbar\kappa\beta)^{3/2}\sum_{v\in V(\gamma)\cap
R}\sqrt{\left|\frac{i}{3\,!\cdot 8}\sum_{e\cap e' \cap
e''=v}\epsilon(e,e',e'')\epsilon_{ijk}X^i_eX^j_{e'}X^k_{e''}\right|}\,\,.
\end{align}
Because the non-vanishing contributions of $\hat{V}$ acting on a
cylindrical function $f_\gamma$ come from the vertices $v\in
V(\gamma)$, only the integration over the tetrahedra
$\Delta^\epsilon_{\gamma,v,e_I,e_J, e_K}$ needs to be considered.
Hence the resulted operator corresponding the last two integrals in
Eq. (\ref{regE1}) acts on a cylindrical function as
\begin{align}
\label{int1} &\frac{1}{36i^6\hbar^6}\sum_{v'\in
V(\gamma)}\chi^3_\epsilon(x,v')\frac{8}{E(v')}\sum_{v(\Delta)=v'}\epsilon(s_Is_Js_K)\epsilon^{IJK}
\mathrm{Tr}\Big(h_{s_I(\Delta)}\left[h_{s_I(\Delta)}^{-1},\sqrt{\hat{V}(v',\epsilon)}\;\right]\nonumber\\
&\quad\quad\quad\quad\quad\quad\times
   h_{s_J(\Delta)}\left[h_{s_J(\Delta)}^{-1},\sqrt{\hat{V}(v',\epsilon)}\;\right]
   h_{s_K(\Delta)}\left[h_{s_K(\Delta)}^{-1},\sqrt{\hat{V}(v',\epsilon)}\;\right]\Big)\nonumber\\
&\quad\times\sum_{v''\in
V(\gamma)}\chi^3_\epsilon(x,v'')\frac{8}{E(v'')}\sum_{v(\Delta')=v''}\epsilon(s_Ls_Ms_N)\epsilon^{LMN}
\mathrm{Tr}\Big(h_{s_L(\Delta')}\left[h_{s_L(\Delta')}^{-1},\sqrt{\hat{V}(v'',\epsilon)}\;\right]\nonumber\\
  &\quad\quad\quad\quad\quad\quad\times
   h_{s_M(\Delta')}\left[h_{s_M(\Delta')}^{-1},\sqrt{\hat{V}(v'',\epsilon)}\;\right]
   h_{s_N(\Delta')}\left[h_{s_N(\Delta')}^{-1},\sqrt{\hat{V}(v'',\epsilon)}\;\right]\Big)\cdot
   f_\gamma.
\end{align}
We now come to the quantization of the second integral in Eq.
(\ref{regE1}). Given a graph $\gamma$ and a 2-surface $S$, we can
change the orientations of some edges of $\gamma$ and subdivide
edges of $\gamma$ into two halves at an interior point if necessary,
and obtain a graph $\gamma_S$ adapted to $S$ such that the edges of
$\gamma_S$ belong to the following four types \cite{ThomasRev}: (i)
$e$ is the up type edge if $e\cap S=b(e)$ and
$\dot{e}^a(0)n_a(e(0))>0$ where $n_a$ is the co-vector field normal
to $S$; (ii) $e$ is the down type edge if $e\cap S=b(e)$ and
$\dot{e}^a(0)n_a(e(0))<0$; (iii) $e$ is the inside type edge if
$e\cap S=e$; (iv) $e$ is the outside type edge if $e\cap
S=\emptyset$. In the following, we only use the graphs adapted to
some 2-surfaces. For convenience, we will abbreviate the coordinate
$(z^1,z^2,y^3)$ of a point in 2-d surface $S_{y^3}$ as $(z,y^3)$. In
a suitable operator-ordering, that integral can be quantized as an
operator acting on a cylindrical function as follows:
\begin{align}
\label{int2} &\int_\Sigma
    d^3y\left[\int_{S_{y^3}}d^2z\frac{\chi^2_{\epsilon'}(y^1,y^2;z^1,z^2)}
   {\hat{Ar}(z,y^3;\epsilon')}\hat{E}^3_i(z^1,z^2,y^3)\right]\left[\partial_a\chi^3_\epsilon(x,y)\right]
   \hat{E}^a_i(y)\cdot f_{\gamma}\nonumber\\
&=\frac{(-i\hbar\kappa\beta)^2}{8}\sum_{e\in E(\gamma)}\int_\Sigma
    d^3y\sum_{e'\in E(\gamma),e'(0)\in
    S_{y^3}}\varrho(e',S_{y^3})\frac{\chi^2_{\epsilon'}(y^1,y^2;\,e'(0))}{\hat{Ar}(e'(0),\epsilon')}
    X^i_{e'}(0)\nonumber\\
 &\quad\quad\times\left[\partial_a\chi^3_\epsilon(x,y)\right]\int_0^1dt\dot{e}^a(t)\delta^3(y,(e(t)))X^i_e(t)\cdot
   f_\gamma\nonumber\\
&=\frac{(-i\hbar\kappa\beta)^2}{8}\sum_{e\in
    E(\gamma)}\lim_{n\to\infty} \sum_{k=1}^n
   [\chi^3_\epsilon(x,e(t_k))-\chi^3_\epsilon(x,e(t_{k-1}))]\nonumber\\
  &\quad\quad\times\sum_{e'\in E(\gamma),e'(0)\in
    {S_{e^3(t_{k-1})}}}\varrho(e',S_{e^3(t_{k-1})})\frac{\chi^2_{\epsilon'}(e^1(t_{k-1}),e^2(t_{k-1});\,e'(0))}
    {\hat{Ar}(e'(0),\epsilon')}X^i_{e'}(0)X^i_e(t_{k-1})\cdot
   f_\gamma,
\end{align}
where $0=t_0<t_1<..<t_n=1$ is an arbitrary partition of the
interval $[0,1]$, $X^i_e(t):= [h_e(0,t)\tau_i h_e(t,1)]_{AB}
\partial/\partial[h_e(0,1)]_{AB}$ (we denote $X^i_e:=X^i_e(0)$ in the
following) and
\begin{equation}
\label{sig} \varrho(e,S)=
\begin{cases}
\,+1,& \text{if $e$ is of the up type with respect to $S$;
      }\\
       -1,& \text{if $e$ is of the down type with respect to $S$}; \\
      0,& \text{if $e$ is of the inside or outside type with respect to $S$}.\nonumber
\end{cases}
\end{equation}
Let us introduce the set of isolated intersection points of $\gamma$
and $S$
\begin{align}
P(\gamma,S):=\{e\cap S| \varrho(e,S)\neq0,e\in E(\gamma)\}.
\end{align}
The
first integral in Eq. (\ref{regE1}) can be quantized
straightforwardly, since $\widehat{\sqrt{\det(\sigma)}}(x)$ is given
by \cite{ALRev,Area}
\begin{align}
\label{int3} \widehat{\sqrt{\det(\sigma)}}(x)\cdot
f_\gamma&=\frac{1}{4}\hbar\kappa\beta\sum_{v\in
P(\gamma,S)}\delta^2(x,v)\sqrt{-\sum_{e,e'\in
E(\gamma);\,b(e)=b(e')=v}\varrho(e,e')X^i_eX^i_{e'}}\cdot f_\gamma\nonumber\\
&=:\sum_{v\in P(\gamma,S)}\delta^2(x,v)\hat{Ar}_v\cdot f_\gamma,
\end{align}
where $\varrho(e,e'):=\varrho(e,S)\varrho(e',S)$. Putting Eqs.
(\ref{int1}), (\ref{int2}) and (\ref{int3}) together, we finally
obtain the regularized operator corresponding to Eq. (\ref{regE1}),
acting on a cylindrical function as
\begin{align}
&\hat{E}^{\epsilon,\epsilon',n}_{Q,k}(S)\cdot f_\gamma\nonumber\\
&=\frac{2^{9}\beta^2}{81\hbar^4\kappa^5}\sum_{v\in
P(\gamma,S)}\hat{Ar}_v\sum_{e\in
    E(\gamma)}\sum_{k=1}^n
   [\chi^3_\epsilon(v,e(t_k))-\chi^3_\epsilon(v,e(t_{k-1}))]\nonumber\\
  &\quad\quad\times\sum_{e'\in E(\gamma),e'(0)\in
    {S_{e^3(t_{k-1})}}}\varrho(e',S_{e^3(t_{k-1})})\frac{\chi^2_{\epsilon'}(e^1(t_{k-1}),e^2(t_{k-1});\,e'(0))}
    {\hat{Ar}(e'(0),\epsilon')}X^i_{e'}(0)X^i_e(t_{k-1})\nonumber\\
  &\quad\quad\times\sum_{v'\in
   V(\gamma)}\chi^3_\epsilon(v,v')\frac{8}{E(v')}\sum_{v(\Delta)=v'}\epsilon(s_Is_Js_K)\epsilon^{IJK}
   \mathrm{Tr}\Big(h_{s_I(\Delta)}\left[h_{s_I(\Delta)}^{-1},\sqrt{\hat{V}(v',\epsilon)}\;\right]\nonumber\\
   &\quad\quad\quad\quad\times h_{s_J(\Delta)}\left[h_{s_J(\Delta)}^{-1},\sqrt{\hat{V}(v',\epsilon)}\;\right]
 h_{s_K(\Delta)}\left[h_{s_K(\Delta)}^{-1},\sqrt{\hat{V}(v',\epsilon)}\;\right]\Big)\nonumber\\
&\quad\times\sum_{v''\in
   V(\gamma)}\chi^3_\epsilon(v,v'')\frac{8}{E(v'')}\sum_{v(\Delta')=v''}\epsilon(s_Ls_Ms_N)\epsilon^{LMN}
   \mathrm{Tr}\Big(h_{s_L(\Delta')}\left[h_{s_L(\Delta')}^{-1},\sqrt{\hat{V}(v'',\epsilon)}\;\right]\nonumber\\
   &\quad\quad\quad\quad\times
  h_{s_M(\Delta')}\left[h_{s_M(\Delta')}^{-1},\sqrt{\hat{V}(v'',\epsilon)}\;\right]
   h_{s_N(\Delta')}\left[h_{s_N(\Delta')}^{-1},\sqrt{\hat{V}(v'',\epsilon)}\;\right]\Big)\cdot
   f_\gamma.
\end{align}
Now we perform the limit $n\rightarrow\infty$,
$\epsilon'\rightarrow 0$ and $\epsilon\rightarrow 0$ in reversed
order. Keeping $n$ fixed, for small enough $\epsilon$, only the
term with $k =1$ in the sum survives provided that $b(e)=v$, and
only terms with $v=v'=v''$ contribute. So for small enough
$\epsilon$, the above operator reduces to
\begin{align}
\hat{E}^{n,\epsilon,\epsilon'}_{Q,k}(S)\cdot f_\gamma
&=-\frac{2^{9}\beta^2}{81\hbar^4\kappa^5}\sum_{v\in
P(\gamma,S)}\hat{Ar}_v\nonumber\\
  &\quad\quad\times\sum_{e'\in E(\gamma),e'(0)\in
    {S_{e^3(t_{0})}}=S}\varrho(e',S_{e^3(t_{0})})\frac{\chi^2_{\epsilon'}(v;\,e'(0))}
    {\hat{Ar}(e'(0),\epsilon')}X^i_{e'}(0)\sum_{b(e)=v}X^i_e(0)\nonumber\\
  &\quad\quad\times\frac{8}{E(v)}\sum_{v(\Delta)=v}\epsilon(s_Is_Js_K)\epsilon^{IJK}\mathrm{Tr}\Big(h_{s_I(\Delta)}
  \left[h_{s_I(\Delta)}^{-1},\sqrt{\hat{V}(v,\epsilon)}\;\right]\nonumber\\
   &\quad\quad\quad\quad\times h_{s_J(\Delta)}\left[h_{s_J(\Delta)}^{-1},\sqrt{\hat{V}(v,\epsilon)}\;\right]
 h_{s_K(\Delta)}\left[h_{s_K(\Delta)}^{-1},\sqrt{\hat{V}(v,\epsilon)}\;\right]\Big)\nonumber\\
&\quad\times\frac{8}{E(v)}\sum_{v(\Delta')=v}\epsilon(s_Ls_Ms_N)\epsilon^{LMN}\mathrm{Tr}\Big(h_{s_L(\Delta')}
\left[h_{s_L(\Delta')}^{-1},\sqrt{\hat{V}(v,\epsilon)}\;\right]\nonumber\\
   &\quad\quad\quad\quad\times
  h_{s_M(\Delta')}\left[h_{s_M(\Delta')}^{-1},\sqrt{\hat{V}(v,\epsilon)}\;\right]
   h_{s_N(\Delta')}\left[h_{s_N(\Delta')}^{-1},\sqrt{\hat{V}(v,\epsilon)}\;\right]\Big)\cdot
   f_\gamma.\nonumber
\end{align}
For small enough $\epsilon'$, the function
$\chi^2_{\epsilon'}(v,e'(0))$ vanishes unless $v=e'(0)$. Hence the
above regularized operator reduces to
\begin{align}
\hat{E}^{n,\epsilon,\epsilon'}_{Q,k}(S)\cdot
f_\gamma&=-\frac{2^{11}\beta^2}{81\hbar^4\kappa^5}\sum_{v\in
P(\gamma,S)}\hat{Ar}_v\frac{1}{\hat{Ar}(v,\epsilon')}\sum_{b(e')=v}\varrho(e',S)
    X^i_{e'}\sum_{b(e)=v}X^i_e\nonumber\\
  &\quad\quad\times\frac{8}{E(v)}\sum_{v(\Delta)=v}\epsilon(s_Is_Js_K)\epsilon^{IJK}\mathrm{Tr}\Big(h_{s_I(\Delta)}
  \left[h_{s_I(\Delta)}^{-1},\sqrt{\hat{V}(v,\epsilon)}\;\right]\nonumber\\
   &\quad\quad\quad\quad\times h_{s_J(\Delta)}\left[h_{s_J(\Delta)}^{-1},\sqrt{\hat{V}(v,\epsilon)}\;\right]
 h_{s_K(\Delta)}\left[h_{s_K(\Delta)}^{-1},\sqrt{\hat{V}(v,\epsilon)}\;\right]\Big)\nonumber\\
&\quad\times\frac{8}{E(v)}\sum_{v(\Delta')=v}\epsilon(s_Ls_Ms_N)\epsilon^{LMN}\mathrm{Tr}\Big(h_{s_L(\Delta')}
\left[h_{s_L(\Delta')}^{-1},\sqrt{\hat{V}(v,\epsilon)}\;\right]\nonumber\\
   &\quad\quad\quad\quad\times
  h_{s_M(\Delta')}\left[h_{s_M(\Delta')}^{-1},\sqrt{\hat{V}(v,\epsilon)}\;\right]
   h_{s_N(\Delta')}\left[h_{s_N(\Delta')}^{-1},\sqrt{\hat{V}(v,\epsilon)}\;\right]\Big)\cdot
   f_\gamma.\nonumber
\end{align}
Notice that $\varrho(e',S)$ implies that the edges inside $S$ have
no contribution to the operation, while the action of area
operator $\hat{Ar}(v,\epsilon')$ on the edges transversal to $S$
is non-vanishing. Hence $1/\hat{Ar}(v,\epsilon')$ is well defined.
Thus one can take the limits and obtain an operator as
\begin{align}
\label{hatE-1}
 \hat{E}_{Q,k}(S)\cdot f_\gamma
&=-\frac{2^{9}\beta^2}{81\hbar^4\kappa^5}\sum_{v\in
P(\gamma,S)}\sum_{b(e')=v}\varrho(e',S)
    X^i_{e'}\sum_{b(e)=v}X^i_e\nonumber\\
  &\quad\times\frac{8}{E(v)}\sum_{v(\Delta)=v}\epsilon(s_Is_Js_K)\epsilon^{IJK}\mathrm{Tr}\Big(h_{s_I(\Delta)}
  \left[h_{s_I(\Delta)}^{-1},\sqrt{\hat{V}_v}\;\right]\nonumber\\
   &\quad\quad\times h_{s_J(\Delta)}\left[h_{s_J(\Delta)}^{-1},\sqrt{\hat{V}_v}\;\right]
 h_{s_K(\Delta)}\left[h_{s_K(\Delta)}^{-1},\sqrt{\hat{V}_v}\;\right]\Big)\nonumber\\
&\quad\times\frac{8}{E(v)}\sum_{v(\Delta')=v}\epsilon(s_Ls_Ms_N)\epsilon^{LMN}\mathrm{Tr}\Big(h_{s_L(\Delta')}
\left[h_{s_L(\Delta')}^{-1},\sqrt{\hat{V}_v}\;\right]\nonumber\\
   &\quad\quad\times
  h_{s_M(\Delta')}\left[h_{s_M(\Delta')}^{-1},\sqrt{\hat{V}_v}\;\right]
   h_{s_N(\Delta')}\left[h_{s_N(\Delta')}^{-1},\sqrt{\hat{V}_v}\;\right]\Big)\cdot
   f_\gamma\nonumber\\
&=:\sum_{v\in P(\gamma,S)}\hat{E}_{Q,k,v}\cdot f_\gamma.
\end{align}
Let $E_{v,*}(\gamma)=\{e\in E(\gamma); v=b(e); e=*\,\mathrm{type}\}$
where $*=\mathrm{u,d,i}$ for `up, down, inside' with respect to $S$
respectively, and let $X_{v,*}^i=\sum_{e\in E_{v,*}}X^i_e$. Then one
can check the commutation relation \cite{ThomasRev}
\begin{align}
[X^i_{v,*},X^j_{v',*'}]=-2\epsilon_{ijk}X^k_{v,*}\delta_{v,v'}\delta_{*,*'}.
\end{align}
Hence one has
\begin{align}
\sum_{b(e')=v}\varrho(e',S)
    X^i_{e'}\sum_{b(e)=v}X^i_e=&(X^i_{v,\mathrm{u}}-X^i_{v,\mathrm{d}})(X^i_{v,\mathrm{u}}+X^i_{v,\mathrm{d}}
    +X^i_{v,\mathrm{i}})\nonumber\\
=&\sum_{b(e)=v}X^i_e\sum_{b(e')=v}\varrho(e',S)
    X^i_{e'}.
\end{align}
Thus there is no operator-ordering problem for these two operators. Moreover, the operator\\
$\sum_{b(e')=v}\varrho(e',S)
    X^i_{e'}\sum_{b(e)=v}X^i_e$ is gauge invariant since
\begin{align}
&\left[\sum_{b(e')=v}\varrho(e',S)
    X^i_{e'}\sum_{b(e)=v}X^i_e,\sum_{b(e'')=v}X^j_{e''}\right]\nonumber\\
=&\left[(X^i_{v,\mathrm{u}}-X^i_{v,\mathrm{d}})(X^i_{v,\mathrm{u}}+X^i_{v,\mathrm{d}}+X^i_{v,\mathrm{i}}),
    X^j_{v,\mathrm{u}}+X^j_{v,\mathrm{d}}+X^j_{v,\mathrm{i}}\right]
= 0.
\end{align}
Therefore our quasi-local energy operator $\hat{E}_{Q,k}(S)$ in
(\ref{hatE-1}) is gauge invariant. Moreover, we can also define a
symmetric quantum version of $E_{Q,k}(S)$ as
\begin{align}
\hat{E}^s_{Q,k}(S):=\frac{1}{2}\left(\hat{E}_{Q,k}(S)+\hat{E}^\dag_{Q,k}(S)\right).
\end{align}
A special property of the $\hat{E}_{Q,k}(S)$ (or
$\hat{E}^s_{Q,k}(S)$) is immediately clear. Because
$\sum_{b(e)=v}X^i_e$ generates the internal gauge transformations,
$\hat{E}_{Q,k}(S)$ (or $\hat{E}^s_{Q,k}(S)$) vanishes on
gauge-invariant states. Note that the quantization of $E_{Q,k}(S)$
is not unique. An alternative quantization with a partial gauge
fixing is given in Appendix \ref{gaugefix}.

\subsection{Quasi-local normal momentum operator}\label{section2}

The so-called normal-directional momentum of $S$ in adapted
coordinates can be expressed as \cite{BLY}
\begin{align}
\label{JQ} J_{Q,l}(S):=&\frac{1}{\kappa}\int_S
d^2x\sqrt{\det(\sigma)}\;l,
\end{align}
where $l$ is the trace of $l_{ab}$, which is the extrinsic
curvature tensor of $S$ with respect to its unit normal $u^a$
orthogonal to $\Sigma$, i.e.,
\begin{align}
l=&\sigma^{ab}l_{ab}=\sigma^{ab}\nabla_au_b.
\end{align}
Let $Ar(S)$ be the area of a closed 2-surface $S$. The Euclidean
Hamiltonian constraint with lapse function $N=1$ reads
\begin{align}
H^E(1)=\frac{1}{2\kappa}\int_\Sigma
d^3x\frac{1}{\sqrt{\det(q)}}\,\epsilon_i^{\ jk}F^i_{ab}E^a_jE^b_k.
\end{align}
Then we have (see Appendix \ref{proof} for a proof)
\begin{eqnarray}
\label{Jident} J_{Q,l}(S)=\frac{1}{\kappa}\big\{H^E(1),Ar(S)\big\}.
\end{eqnarray}
Thus the quasi-local momentum $J_{Q,l}(S)$ of $S$ can be regarded as
the ``time derivative" of the area of $S$. Since there are densely
defined operators corresponding to $H^E(1)$ and $Ar(S)$ in ${\cal
H}_{\mathrm{kin}}$, we may replace Poisson brackets by commutators
times $1/(i\hbar)$, functions by operators, and obtain the operator
$\hat{J}_{Q,l}(S)$. However, there are two quantum versions of
$H^E(1)$ in the literature, the graph-changing one \cite{QSDI,QSDII}
and the no-graph-changing one \cite{AQG}. It turns out that the
no-graph-changing version is more convenient in defining the quantum
version of $\hat{J}_{Q,l}(S)$.

We now introduce the minimal loop prescription \cite{ST}. Given a
vertex $v$ of $\gamma$ and two different edges $e_i, e_j$ incident
at and outgoing from $v$, a loop $\alpha_{ij}$ within $\gamma$
starting at $v$ along $e_i$ and ending at $v$ along $e_j^{-1}$ is
said to be minimal provided that there is no other loop within
$\alpha_{ij}$ satisfying the same restrictions with fewer edges
traversed. We denote by $L(v,e_i,e_j)$ the set of minimal loops with
the data indicated. The non-graph-changing symmetric operator
$\hat{H}^E(1)$ acts on a cylindrical function as \cite{AQG}
\begin{align}
\hat{H}^E(1)\cdot f_\gamma&=\frac{1}{3i\hbar\kappa^2\beta}\sum_{v\in
V(\gamma)}\sum_{e_i\cap e_j\cap
e_k=v}\epsilon^{ijk}\frac{\epsilon(e_i,e_j,e_k)}{|L(v,e_i,e_j)|}\nonumber\\
&\quad\quad\quad\quad\times\sum_{\alpha_{ij}\in
L(v,e_i,e_j)}\mathrm{Tr}\left(\left\{h_{\alpha_{ij}},h_{e_k}[h^{-1}_{e_k},\hat{V}_v]\right\}\right)
\cdot f_\gamma\nonumber\\
&=:\sum_{v\in V(\gamma)}\hat{H}^E_v\cdot f_\gamma,
\end{align}
where $\{\cdot,\cdot\}$ denotes the anti-commutator.

Hence it is easy to obtain a well-defined operator corresponding to
$J_{Q,l}(S)$ as
\begin{eqnarray}
\label{J_Ql}
\hat{J}_{Q,l}(S)=\frac{1}{i\hbar\kappa}\big[\hat{H}^E(1),\hat{Ar}(S)\big].
\end{eqnarray}
It is easy to show that $\hat{J}_{Q,l}(S)$ is a gauge-invariant,
diffeomorphism-covariant and symmetric operator. For later purposes
we write a more explicit expression for $\hat{J}_{Q,l}(S)$ operating
on a cylindrical function as
\begin{align}
\hat{J}_{Q,l}(S)\cdot f_\gamma&=\frac{1}{i\hbar\kappa}\sum_{v'\in
V(\gamma),v\in P(\gamma,S)}\big[\hat{H}^E_{v'},\hat{Ar}_v\big]\cdot
f_\gamma=\frac{1}{i\hbar\kappa}\sum_{v\in
P(\gamma,S)}[\hat{H}^E_v,\hat{Ar}_v]\cdot f_\gamma\nonumber\\
&=:\sum_{v\in P(\gamma,S)}\hat{J}_{Q,l,v}\cdot f_\gamma,
\end{align}
where in the first step we used the fact that $\hat{H}^E(1)$ is a
non-graph-changing operator, and in the second step we exploited
that $\hat{Ar}_v$ only acts on the edges incident at $v\in
P(\gamma,S)$ so that the commutator with $\hat{H}^E_{v'}$, which
contains only holonomies of edges incident at $v'$, vanishes if
$v'\neq v$.

\section{QLE operators}

In this section, we will use the two well-defined operators
$\hat{E}_{Q,k}(S)$ and $\hat{J}_{Q,l}(S)$ constructed in the last
section as building blocks to quantize several types of QLE
expressions.

\subsection{Brown-York energy operator}

The system under consideration is a spatial three-surface $\Sigma$
bounded by a two-surface $S$ in a spacetime region that can be
decomposed as a product of a spatial three-surface and a real
line-interval representing time. Suppose that the 2-metric
$\sigma_{ab}$ induced on $S$ has positive scalar curvature. Then
by the embedding theorem there is a unique isometric embedding of
$(S,\sigma_{ab})$ into the flat 3-space. Let $k_o$ be the trace of
extrinsic curvature of $S$ in this embedding, which is completely
determined by $\sigma_{ab}$ and is necessarily positive.  The time
evolution of the two-surface boundary $S$ is the timelike
three-surface boundary ${}^3\!B$. Brown and York defined their QLE
by the Hamiltonian-Jacobi method as \cite{Brown-York}:
\begin{align}
\label{0} E_{BY}(S):=\frac{1}{\kappa}\int_S
d^2x\sqrt{\det(\sigma)}(k_o-k),
\end{align}
where $k$ is the trace of the extrinsic curvature $k_{ab}$ of $S$
corresponding to the normal $n^a$ orthogonal to $ {}^3\!B$
(``orthogonal boundaries assumption"), and the integral of ${k_o}$
is a reference term that is used to normalize the energy with
respect to a reference spacetime, not necessarily flat. The second
integral in Eq. (\ref{0}) have been quantized as Eq. (\ref{hatE-1}).
The construction of reference term from a reference space amounts to
posing and solving an isomeric embedding problem. One natural choice
is to embed $S$ isometrically into Euclidean three space
$(\mathbb{R}^3, \delta_{ab})$ in order to obtain an extrinsic
curvature tensor $(k_o)_{ab}$ (and hence $k_o$). With the second
fundamental form $(k_o)_{ab}$ expressed in terms of the embedding's
coordinate chart, this is the Weyl's problem, a classic problem of
differential geometry for which an extensive literature exists
\cite{Weyl}. However, it is very difficult to obtain the solution of
$(k_o)_{ab}$ in terms of $\sigma_{ab}$. Nevertheless, since the
function of the reference term is to normalize the energy, in the
quantum version one may regard it as a c-number
$-K_o\equiv\frac{1}{\kappa}\int_S d^2x\sqrt{\det(\sigma)}\,k_o$. In
this sense the Brown-York energy has been quantized as
\begin{align}
\hat{E}_{BY}(S)=\hat{E}_{Q,k}(S)-K_o.
\end{align}
Of course, one may also take the other viewpoint that the
reference term is dynamical and thus should be quantized. In
certain symmetric models, it is indeed possible to solve $k_o$.

\subsubsection{QLE in spherically symmetric model}

We now study the Brown-York QLE in spherically symmetric quantum
geometry. Consider a static spherically symmetric space-time with
line element
\begin{align}
\label{line} ds^2=-N^2dt^2+H^2dr^2+R^2(d\theta^2+\sin^2\theta
d\phi^2),
\end{align}
where $N$ and $H$ are functions of $r$ only. Let $\Sigma_t$ be the
interior of a $t=constant$ slice with two-boundary
$S_{r=\rm{const}}$ specified by $r=constant$. A straightforward
calculation gives the trace $k$ of $k_{ab}$ as
\begin{align}
k=\frac{2R'}{RH}\,,
\end{align}
where the prime denotes the partial differentiation with respect
to coordinate $r$. Now consider a round sphere with radius $R$
embedded in $(\mathbb{R}^3,\delta_{ab})$. Such a sphere has an
extrinsic curvature $(k_o)_{ab}$ with trace
\begin{align}
k_o=\frac{2}{R}\,.
\end{align}
So the Brown-York quasi-local energy can be written as
\begin{align}
E_{BY}(S_{r=\rm{const}})
         &=\frac{1}{\kappa}\int_0^{2\pi}d\varphi\int_0^\pi d\theta
            R^2\sin\theta\left(\frac{2}{R}-\frac{2R'}{RH}\right)\nonumber\\
         &=\frac{8\pi}{\kappa}R(1-\frac{R'}{H}).
\end{align}
In the spherically symmetric model, the invariant connections and
triads can be written as \cite{SphereRed}
\begin{align}
 A_a=&A_r(r)\Lam_3(\md r)_a+\left[A_1(r)\Lam_1+A_2(r)\Lam_2\right](\md\theta)_a+
\left[A_1(r)\Lam_2-A_2(r)\Lam_1\right]\sin\theta(\md\vp)_a\nonumber\\
&+\Lam_3\cos\theta(\md\vp)_a,\nonumber\\
E^a=&E^r(r)\Lam_3\sin\theta\left(\frac{\partial}{\partial
r}\right)^a+
\left[E^1(r)\Lam_1+E^2(r)\Lam_2\right]\sin\theta\left(\frac{\partial}{\partial\theta}\right)^a\nonumber\\
&+\left[E^1(r)\Lam_2-E^2(r)\Lam_1\right]\left(\frac{\partial}{\partial\vp}\right)^a,
\end{align}
where $A_r,A_1,A_2,E^r,E^1$ and $E^2$ are real functions on an
one-dimensional, radial manifold $M$ with coordinate $r$, and
$\Lambda_I$ are the $su(2)$-matrices and identical to $\tau_i/2$
or a rigid rotation thereof. For convenience, one introduces
variables
\begin{align}
A_\varphi(r)&:=\sqrt{(A_1(r))^2+(A_2(r))^2},\\
E^\varphi(r)&:=\sqrt{(E^1(r))^2+(E^2(r))^2},
\end{align}
and $\alpha(r)$, $\beta(r)$ defined by
\begin{align}
\Lambda_1\cos\beta(r)+\Lambda_2\sin\beta(r)&=\left(A_1(r)\Lambda_2-A_2(r)\Lambda_1\right)/A_\varphi(r),\\
\Lambda_1\cos\left(\alpha(r)+\beta(r)\right)+\Lambda_2\sin\left(\alpha(r)+\beta(r)\right)&=
\left(E^1(r)\Lambda_2-E^2(r)\Lambda_1\right)/E^\varphi(r).
\end{align}
In the spherical coordinate system, the components of the spatial
3-metric $q_{ab}$ on $\Sigma$ take the form
\begin{align}
(q_{ab})=\mathrm{diag}\left(\frac{(E^\varphi)^2}{E^r},E^r,E^r\sin^2\theta\right).
\end{align}
So we have the relation
\begin{align}
H&=\frac{E^\varphi}{\sqrt{E^r}}\,\, ,
\;\;\;\;\;\;\;\;\;R=\sqrt{E^r}.
\end{align}
The QLE reads
\begin{align}
\label{ES}
E_{BY}(S_{r=\rm{const}})&=\frac{8\pi}{\kappa}\sqrt{E^r}\left[1-\frac{(E^r)'}{2E^\varphi}\right].
\end{align}
In Ref. \cite{SphereH}, a canonical transformation from
$(A_r,A_1,A_2;E^r,E^1,E^2)$ to $(A_r,\beta
K_\varphi,\eta;E^r,E^\varphi,P^\eta)$, where
\begin{align}
K_\varphi(r)&:=\sqrt{(K_1(r))^2+(K_2(r))^2}
\end{align}
is the extrinsic curvature component and
\begin{align}
\eta(r)&:=\alpha(r)+\beta(r),\\
P^\eta(r)&:=2A_\varphi(r)E^\varphi(r)\sin\alpha(r)
\end{align}
has been made. The symplectic structure is then given by
\begin{align}
\Omega_B
 &=\frac{4\pi}{\kappa\beta}\int_Bdr(\delta
    A_r\wedge \delta E^r+2\beta\delta K_\varphi\wedge\delta E^\varphi+\delta\eta\wedge\delta
    P^\eta).
\end{align}
The resulted Hilbert space is spanned by an orthonormal basis of
spin network states:
\begin{equation} \label{SpinNetwork}
 T_{g,k,\mu}(A) = \prod_{e\in g}
 \exp\left(\tfrac{1}{2}ik_e\smallint_e A_r(r)\md r\right) \prod_{v\in
 V(g)} \exp(i\mu_v \beta K_{\vp}(v)) \exp(ik_v\eta(v))
\end{equation}
with edge labels $k_e\in\Z$, vertex labels $\mu_v\in\R$ and
$k_v\in\Z$ for graphs $g$ in the 1-dimensional radial manifold
$M$. The flux operators corresponding to the momenta $E^r$ and
$E^\varphi$ act on the spin network states as:
\begin{align}
\hat{E}^r(r)T_{g,k,\mu}&=\frac{\beta\hbar\kappa}{8\pi}\frac{k_{e^+(r)}+k_{e^-(r)}}{2}T_{g,k,\mu},\\
\int_{\cal
I}\hat{E}^\varphi(r)T_{g,k,\mu}&=\frac{\beta\hbar\kappa}{4\pi}\sum_{v\in
{\cal I}}\mu_v\delta(r,v)T_{g,k,\mu},
\end{align}
where $e^{\pm}(r)$ are the two edges (or two parts of a single
edges) meeting at $r$, $\delta(r,v)$ is the Dirac delta
distribution, and ${\cal I}$ is a region of the reduced (radial)
manifold $M$.

We now consider the quantization of the QLE. The spin connection
component can be regularized as \cite{SphereH}
\begin{align}
\label{sphe1} -\frac{(E^r)'}{2E^\varphi}
&=-\frac{1}{4}\left(\frac{E^r(v_+)-E^r(v)}{\int_v^{v_+}E^\varphi
dr}+\frac{E^r(v)-E^r(v_-)}{\int_{v_-}^vE^\varphi
dr}\right)+O(\epsilon).
\end{align}
Now the $E^r$ in Eq. (\ref{sphe1}) can be promoted as well-defined
operators, and the $\frac{1}{\int E^\phi dr}$ may also become
well-defined operator by suitable treatments \cite{Amb,Bojcosmol}.
Hence the quasi-local energy operator
$\hat{E}_{BY}(S_{r=\rm{const}})$ in the spherically symmetric sector
of LQG can be well defined.

\subsection{Liu-Yau energy operator}

Let $l$ be the trace of the extrinsic curvatures of the 2-surface
$S$ in the physical spacetime corresponding to the future pointing
timelike normal. Liu and Yau define a quasi-local energy by
\cite{Liu-Yau}
\begin{eqnarray}
\label{E_LY}
E_{LY}(S):=\frac{1}{\kappa}\int_Sd^2x\sqrt{\det(\sigma)}(k_o-\sqrt{|k^2-l^2|}\,).
\end{eqnarray}
Since the first term in Eq. (\ref{E_LY}) is again a reference term,
we will only consider the second term
\begin{eqnarray}
\label{E_Qkls}
E_{Q,k,l}(S)\equiv-\frac{1}{\kappa}\int_Sd^2x\sqrt{\det(\sigma)}\sqrt{|k^2-l^2|}.
\end{eqnarray}
Let $g_\epsilon(x,y)$ be a 1-parameter family of fields on S which
tend to $\delta^2(x,y)$ as $\epsilon$ tends to zero, i.e., such
that
\begin{eqnarray}
\label{geps} \lim_{\epsilon\rightarrow
0}\int_Sd^2y\,g_\epsilon(x^1, x^2; y^1, y^2)f(y^1, y^2)=f(x^1,
x^2)
\end{eqnarray}
for all smooth densities $f$ of weight 1 and of compact support on
$S$. (Thus, $g_\epsilon(x,y)$ is a density of weight 1 in $x$ and
a function in $y$.) Using $g_\epsilon(x,y)$ as smearing function,
one can regularize $E_{Q,k,l}(S)$ as
\begin{align}
\label{E_Qkl} E_{Q,k,l}^\epsilon(S)
=&-\int_Sd^2x\left[\Big|\frac{1}{\kappa}\int_Sd^2y\,g_\epsilon(x,y)\sqrt{\det(\sigma)}(-k+l)\,
\frac{1}{\kappa}\int_Sd^2zg_\epsilon(x,z)\sqrt{\det(\sigma)}(-k-l)\Big|\right]^{1/2}\nonumber\\
\equiv&-\int_Sd^2x\left[\Big|\big[E_{Q,-k+l}\big]_{g_{\epsilon}}(x)
\big[E_{Q,-k-l}\big]_{g_{\epsilon}}(x)\Big|\right]^{1/2}
\equiv-\int_Sd^2x\,\sqrt{[E_S]_{g_{\epsilon}}}(x).
\end{align}
It is easy to see that $E_{Q,k,l}^\epsilon(S)$ tends to
$E_{Q,k,l}(S)$ as $\epsilon$ tends to zero. Let us now turn to the
integrand of Eq. (\ref{E_Qkl}). It can be promoted as an operator
acting on a cylindrical function as
\begin{align}
[\hat{E}_S]_{g_{\epsilon}}(x)\cdot f_\gamma=&\left|[\hat{E}_{Q,-k+l}]_{g_{\epsilon}}(x)
[\hat{E}_{Q,-k-l}]_{g_{\epsilon}}(x)\right|\cdot f_\gamma\nonumber\\
=&\sum_{v,v'\in
P(\gamma,S)}g_\epsilon(x,v)g_\epsilon(x,v')\big|(\hat{E}_{Q,k,v}+\hat{J}_{Q,l,v})(\hat{E}_{Q,k,v'}
-\hat{J}_{Q,l,v'})\big|\cdot f_\gamma,
\end{align}
where the absolute value
$\big|(\hat{E}_{Q,k,v}+\hat{J}_{Q,l,v})(\hat{E}_{Q,k,v}-\hat{J}_{Q,l,v})\big|\equiv\hat{A}$
indicates that one is supposed to take the square root of the
operator $\hat{A}^\dag\hat{A}$. We choose $\epsilon$ sufficiently
small so that $g_\epsilon(x,v)g_\epsilon(x,v')$ is zero unless
$v=v'$. Then one obtains
\begin{align}
[\hat{E}_S]_{g_{\epsilon}}(x)\cdot f_\gamma=&\sum_{v\in
P(\gamma,S)}{g_{\epsilon}}(x,v)^2\big|(\hat{E}_{Q,k,v}+\hat{J}_{Q,l,v})(\hat{E}_{Q,k,v}-\hat{J}_{Q,l,v})\big|\cdot
f_\gamma.
\end{align}
Notice that $[\hat{E}_S]_{g_{\epsilon}}(x)$ is a non-negative
self-adjoint operator and hence have a well defined square root,
which is also a positive-definite self-adjoint operator. Since we
have chosen $\epsilon$ to be sufficiently small, for any given
point $x$ in $S$, $g_\epsilon(x,v)$ is non-zero for at most one
vertex $v$. We can therefore take the sum over $v$ outside the
square root and thus obtain
\begin{align}
\label{Ely1} \sqrt{[\hat{E}_S]_{g_{\epsilon}}}(x)\cdot
f_\gamma=\sum_{v\in
P(\gamma,S)}g_\epsilon(x,v)\sqrt{\big|(\hat{E}_{Q,k,v}+\hat{J}_{Q,l,v})(\hat{E}_{Q,k,v}-\hat{J}_{Q,l,v})\big|}\cdot
f_\gamma.
\end{align}
Finally, we can remove the regulator. By integrating both sides of
Eq. (\ref{Ely1}) on $S$ and then taking the limit
$\epsilon\rightarrow 0$, we obtain the desired operator
corresponding to Eq. (\ref{E_Qkls}) as
\begin{align}
\hat{E}_{Q,k,l}(S)\cdot f_\gamma=-\sum_{v\in
P(\gamma,S)}\sqrt{\left|(\hat{E}_{Q,k,v}+\hat{J}_{Q,l,v})(\hat{E}_{Q,k,v}-\hat{J}_{Q,l,v})\right|}\cdot
f_\gamma.
\end{align}

\subsection{Hawking energy operator}

By studying the perturbation of the dust-filled open
Friedmann-Robertson-Walker space-time, Hawking found that
\cite{Hawking}
\begin{eqnarray}
E_H(S):=\frac{2\sqrt{\pi}}{\kappa}\sqrt{Ar(S)}\left[1-\frac{1}{16\pi}\int_Sd^2x\sqrt{\det(\sigma)}
\;(k^2-l^2)\right]
\end{eqnarray}
behaves as an appropriate notion of energy surrounded by the
space-like topological 2-sphere $S$. The virtue of Hawking energy
is that it does not need a reference term. We can regularize it as
\begin{align}
\label{mHak} E_H(S)
=&\frac{2\sqrt{\pi}}{\kappa}\sqrt{Ar(S)}-\lim_{\epsilon\rightarrow
0}\frac{\sqrt{Ar(S)}}{8\sqrt{\pi}\kappa}\int_Sd^2x\sqrt{\det(\sigma)}(-k+l)\nonumber\\
&\times\int_Sd^2y\frac{\chi^2_\epsilon(x,y)}{\epsilon^2\sqrt{\det(\sigma)}}\sqrt{\det(\sigma)}(-k-l)\nonumber\\
=&\frac{2\sqrt{\pi}}{\kappa}\sqrt{Ar(S)}-\lim_{\epsilon\rightarrow
0}\frac{\sqrt{Ar(S)}}{8\sqrt{\pi}\kappa}\int_Sd^2x\sqrt{\det(\sigma)}(-k+l)\nonumber\\
&\times\int_Sd^2y\frac{\chi^2_\epsilon(x,y)}{Ar(x,\epsilon)}\sqrt{\det(\sigma)}(-k-l).
\end{align}
To quantize the expression (\ref{mHak}), one can replace the
$\sqrt{Ar(S)}$ by $\sqrt{\hat{Ar}(S)}$ and use the well-defined
operators $\hat{E}_{Q,k,v}$ and $\hat{J}_{Q,l,v}$. We then
formally get an operator acts on cylindrical functions as
\begin{align}
\hat{E}^\epsilon_H(S)\cdot
f_{\gamma}=&\frac{2\sqrt{\pi}}{\kappa}\sqrt{\hat{Ar}(S)}\cdot
f_{\gamma}-\frac{1}{8\sqrt{\pi}\kappa}\sum_{v\in P(\gamma,S)}\left(\hat{E}_{Q,k,v}+\hat{J}_{Q,l,v}\right)
\frac{\sqrt{\hat{Ar}(S)}}{\hat{Ar}_v}\times\nonumber\\
&\quad\quad\quad\quad\quad\quad\quad\quad\quad\quad\times\sum_{v'\in
P(\gamma,S)}\chi^2_\epsilon(v,v')(\hat{E}_{Q,k,v'}-\hat{J}_{Q,l,v'})\cdot
f_{\gamma}.
\end{align}
For sufficiently small $\epsilon$, $\chi^2_\epsilon(v,v')$ is zero
unless $v=v'$. Thus one has
\begin{align}
&\hat{E}_H(S)\cdot f_{\gamma}\nonumber\\
=&\frac{2\sqrt{\pi}}{\kappa}\sqrt{\hat{Ar}(S)}\cdot
f_{\gamma}-\frac{1}{8\sqrt{\pi}\kappa}\sum_{v\in
P(\gamma,S)}(\hat{E}_{Q,k,v}+\hat{J}_{Q,l,v})\frac{\sqrt{\hat{Ar}(S)}}{\hat{Ar}_v}(\hat{E}_{Q,k,v}
-\hat{J}_{Q,l,v})\cdot
f_{\gamma}.
\end{align}
However, this is not a densely defined operator due to
$\frac{1}{\hat{Ar}_v}$. Fortunately, since $\hat{E}_{Q,k,v}$
vanishes the internal gauge-invariant states, for a
gauge-invariant state $\Psi_{\gamma}\in {\cal H}_o$ we have
\begin{align}
\hat{E}_H(S)\cdot\Psi_{\gamma}=\frac{2\sqrt{\pi}}{\kappa}\sqrt{\hat{Ar}(S)}\cdot\Psi_{\gamma}
+\frac{1}{8\sqrt{\pi}\kappa}\sum_{v\in
P(\gamma,S)}\hat{J}_{Q,l,v}\frac{\sqrt{\hat{Ar}(S)}}{\hat{Ar}_v}\,\hat{J}_{Q,l,v}\cdot\Psi_\gamma.
\end{align}
Hence we can obtain a well-defined operator corresponding to
Hawking energy in ${\cal H}_o$ as
\begin{align}
\hat{E}_H(S)\cdot\Psi_{\gamma}=&\frac{2\sqrt{\pi}}{\kappa}\sqrt{\hat{Ar}(S)}\cdot\Psi_{\gamma}\nonumber\\
&+\frac{1}{2\sqrt{\pi}\,\hbar^2\kappa^3}\sum_{v\in
P(\gamma,S)}\left[\hat{H}^E_v,\sqrt{\hat{Ar}_v}\,\right]^\dag\sqrt{\hat{Ar}(S)}\,
\left[\hat{H}^E_v,\sqrt{\hat{Ar}_v}\,\right]\cdot\Psi_{\gamma}.
\end{align}
It is easy to see that this operator is symmetric in ${\cal H}_o$.

\subsection{Geroch energy operator}

Geroch modified the Hawking energy and gave the other definition
for QLE as \cite{GE}
\begin{eqnarray}
\label{ECG} E_G(S):=\sqrt{\frac{Ar(S)}{16\pi
G^2}}\,\left(1-\frac{1}{16\pi}\int_Sd^2x\sqrt{\det(\sigma)}\,k^2\right).
\end{eqnarray}
It can be regularized as
\begin{align}
E_G(S)=&\frac{2\sqrt{\pi}}{\kappa}\sqrt{Ar(S)}-\lim_{\epsilon\rightarrow
0}\frac{\sqrt{Ar(S)}}{8\sqrt{\pi}\kappa}\int_Sd^2x\frac{\sqrt{\det(\sigma)}}{Ar(x,\epsilon)}\,
k\int_Sd^2y\chi^2_\epsilon(x,y)\sqrt{\det(\sigma)}\,k.
\end{align}
The quantum operator corresponding to the Geroch energy then
formally reads
\begin{align}
\hat{E}_G(S)\cdot
f_\gamma(A)=&\frac{2\sqrt{\pi}}{\kappa}\sqrt{\hat{Ar}(S)}\cdot
f_\gamma(A)-\frac{1}{8\sqrt{\pi}\kappa}\sqrt{\hat{Ar}(S)}\sum_{v\in
P(\gamma,S)}\frac{1}{\hat{Ar}_v}\hat{E}_{Q,k,v}^2\cdot f_\gamma(A).
\end{align}
Let $T_s$ be the gauge-invariant spin network function in ${\cal
H}_o$. Then $\hat{E}_G(S)$ acts on $T_s$ as
\begin{align}
\label{E_G} \hat{E}_G(S)\cdot T_s
=\frac{2\sqrt{\pi}}{\kappa}\sqrt{\hat{Ar}(S)}\cdot
T_s=\frac{2\sqrt{\pi}}{\kappa}\sqrt{\sum_{v\in
P(\gamma,S)}\hat{Ar}_v}\cdot T_s.
\end{align}
Thus $\hat{E}_G(S)$ is well defined in ${\cal H}_o$. Hence the
spectrum of the area operator \cite{Area} implies the spectrum of
$\hat{E}_G(S)$ as
\begin{align}
\mathrm{Spec}[\hat{E}_G(S)]=\sqrt{\frac{2\,\pi\hbar\beta}{\kappa}}\left[\sum_{v\in
P(\gamma,S)}\sqrt{2j^{(d)}_v(j^{(d)}_v+1)+2j^{(u)}_v(j^{(u)}_v+1)-j^{(d+u)}_v(j^{(d+u)}_v+1)}\,\right]^{\frac{1}{2}},
\end{align}
where $j^{(d)}_v$, $j^{(u)}_v$ and $j^{(d+u)}_v$ are half-integers
subject to the usual condition:
\begin{align}
j^{(d+u)}_v\in\left\{|j^{(d)}_v-j^{(u)}_v|,
|j^{(d)}_v-j^{(u)}_v|+1,\dots,j^{(d)}_v+j^{(u)}_v\right\}.
\end{align}
In the case that all edges of the graph $\gamma$ underlying $T_s$
puncture $S$, i.e., $\gamma$ has no edges tangential to $S$, the
spectrum is reduced to
\begin{align}
\label{SpecE} \mathrm{Spec}[\hat{E}_G(S)]
=&\sqrt{\frac{4\,\pi\hbar\beta}{\kappa}}\left[\sum_{v\in
P(\gamma,S)}\sqrt{j_v(j_v+1)}\right]^{\frac{1}{2}}=:m\left[\sum_{v\in
P(\gamma,S)}\sqrt{j_v(j_v+1)}\,\right]^{\frac{1}{2}},
\end{align}
where $j_v$ are half-integers. In summary, $\hat{E}_G(S)$ is a
densely defined, positive semi-definite operator in ${\cal H}_o$,
and its spectrum is entirely discrete inherited from the property of
area operator. Thus we have proved a quantum positivity QLE theorem.
Moreover, $\hat{E}_G(S)$ is both internal gauge invariant and
invariant under the diffeomorphism transformations tangent to $S$.
Furthermore, the Geroch gravitational energy in a quantum state
labelled by a graph $\gamma$ is concentrated at the vertices of
$\gamma$ which live on 2-surface $S$ and the edges which puncture
$S$ transversely. The discreteness of quantum gravitational energy
enable us to estimate the statistical entropy of the region enclosed
by $S$ in next section.

\section{Discussions: entropy-area relation in LQG}

It was first speculated by Bekenstein that one could associate an
entropy $\mathbb{S}_{BH}$ to a black hole with horizon area $A$ as
\cite{Beken}
\begin{align}
\label{entr-ar} \mathbb{S}_{BH}=k_B \frac{A}{4\hbar G}\,,
\end{align}
where $k_B$ is the Boltzmann constant. The statistically mechanical
origin of this entropy has been an outstanding mystery for
physicists. Some intuitive arguments and accurate calculations have
been done in the framework of LQG to account for Eq.(\ref{entr-ar})
\cite{Rove,AshBH,DL,GM}. Now we have the quantum gravitational
energy of any finite region bounded by a closed 2-surface $S$. So,
in principle, one may study the thermodynamical properties of an
arbitrary bounded gravitational system in LQG by the standard
statistical mechanics method. For example, if one considers it as a
canonical system, the partition function reads
\begin{align}
Z(S) = \mathrm{Tr}\ e^{-\frac{\hat{E}(S)}{k_BT}}
\end{align}
for certain QLE operator $\hat{E}(S)$. Then all the
thermodynamical quantities including the entropy of the system can
be derived in principle. However, since the spectrum of the above
QLE operators are either too complicated or unknown yet, it is
still difficult to do practical calculations. Thus, further
investigations in this strict approach are needed to understand
the spectrum properties of the QLE operators.

On the other hand, the so-called holographic principle says that,
at the fundamental (quantum) level, one should be able to
characterize the state of any physical system located in a bounded
spatial domain by degrees of freedom on the surface of the domain.
Consequently, the number of physical degrees of freedom in the
domain is bounded from above by the area of the boundary of the
domain instead of its volume. If the entropy representing the
degrees of freedom including gravity in a bounded domain could be
calculated, one would be able to check whether the holographic
principle is valid or not in LQG. Some QLE operator and the
corresponding partition function provide a possible approach to
this issue. Again, we need more control on the spectrum of the QLE
operators constructed above.

Nevertheless, the virtue of Geroch QLE operator $\hat{E}_G(S)$ in
(\ref{E_G}) is beneficial for us to generalize the entropy-area
relation in the framework of LQG. Our discussion is restricted a
simple self-gravitating system bounded by a space-like closed
2-dimensional surface $S$. One can count the number of quantum
states corresponding to eigenvalue of $\hat{E}_G(S)$. Thus, it is
regarded as a microcanonical ensemble where the energy of the system
is fixed. The spectrum (\ref{SpecE}) of the Geroch energy operator
implies that the energy eigenvalue involves only the number $N$ of
punctures and the spins $\vec{j}_v$ of the edges that intersect the
surface $S$. Thus the number of the eignstates of a given eignvalue
of $E_G(S)$ is infinite because different positions of the punctures
give different states. This is no longer the case after modeling out
the spatial diffeomeorphisms tangent to $S$. Following \cite{Rove,
Krasnov}, we shall treat the punctures as distinguishable. Our task
is to count the number ${\cal{N}}(E_G(S))$ of quantum states
corresponding to the classical QLE $E_G(S)$. It is to see that
${\cal{N}}(E_G(S))$ is the same as the number ${\cal{N}}(Ar(S))$
corresponding to the classical area
$Ar(S)=\frac{\kappa^2}{4\pi}E_G(S)^2$. Hence the states which we are
considering satisfy
\begin{align}
Ar(S)=\kappa\hbar\beta\sum_jn_j\sqrt{j(j+1)},
\end{align}
where $n_j$ is the number of punctures with spin $j$. Following
\cite{GM}, the number of the states is given by
\begin{align}
{\cal
N}(Ar(S))=\frac{(\sum_jn_j)!}{\prod_jn_j!}\prod_j(2j+1)^{n_j}\,.
\end{align}
Using Stirling's formula, one gets the entropy as \cite{GM}
\begin{align}
\mathbb{S}=k_B\ln {\cal N}(E_G(S))=k_B\ln {\cal
N}(Ar(S))=k_B\frac{Ar(S)}{4\hbar G}\frac{\beta_0}{\beta}\,,
\end{align}
where $\beta_0$ is the solution of the equation
\begin{align}
1=\sum_j(2j+1)e^{-2\pi\beta_0\sqrt{j(j+1)}}\,.
\end{align}
Moreover, if one assumes that ${\cal N}(E_G(S))$ represent the
physical degrees of freedom in the domain bounded by $S$, the
holographic principle is realized in LQG. Note that for marginally
trapped surfaces, the QLE must be the irreducible mass
$\sqrt{A(S)/16\pi G^2}$ \cite{QLEreview}. Hence our discussion for
these cases do not depend on a specific definition of QLE.

We conclude with a few open issues related to the present work.
(i) More understanding of the spectrum of the QLE operators is
needed in order to do further practical calculations. (ii)
Semiclassical analysis on the QLE operators is yet to be done.
(iii) Since classically the integral of $k$ is non-zero in
general, the vanishing of QLE-like operator $\hat{E}_{Q,k}(S)$ on
gauge invariant states is in some sense awkward. One may consider
other possible operator orderings in its construction in order to
avoid the weakness. (iv) The knowledge on the physical Hilbert
space of LQG would be a great help to our scheme.

\section*{Acknowledgements}

We would like to thank Thomas Thiemann for helpful discussion.
This work is a part of project 10675019 supported by NSFC.

\section*{Appendix}

\begin{appendix}
\section{The partial gauge
fixing version of $\hat{E}_{Q,k}(S)$}\label{gaugefix}

Classically,
we have
\begin{align}
k=&D_an^a=D_a\left(\frac{1}{\sqrt{\det(q)}}n^iE^a_i\right)
 =\frac{1}{\sqrt{\det(q)}}\partial_a(n^iE^a_i),
\end{align}
where we have introduced the internal vector $n^i=n_ae^a_i$ in the
``internal space". It is convenient to first carry out a partial
gauge fixing. Let us fix an internal vector field $n^i$ with
$n^in_i=1$. We restrict ourselves to flat derivative operators
$\partial$ which annihilate $n^i$ in addition to $\delta_{ij}$. We
call the partial gauge fixing as $n^i$-gauge fixing. Fixed
$\vec{n}$ gives a fixed direction in the internal gauge group
$SU(2)$. Thus the structure group is reduced from $SU(2)$ to
$U(1)$. Physical states and observables should, of course, be
independent of this choice. The $n^i$ gauge transformations on the
2-d surface $S$, which keep $n^i$ invariant, are generated by the
following Gauss constraint
\begin{align}
\label{Gauss} G(\lambda n^i)=\int d^3x\,\lambda\,n^iG_i,
\end{align}
where $\lambda$ is an arbitrary real number. Hence the
corresponding Gauss constraint operator is given by
\begin{align}
\hat{G}(\lambda n^i)\cdot
f_\gamma=-\frac{i\hbar\kappa\beta}{2}\,\lambda\,n^i\sum_{e\in
E(\gamma)}X^i_e\cdot f_\gamma.
\end{align}
Let $\psi_\gamma$ be a $n^i$-gauge-invariant cylindrical function
corresponding the above Gauss constraint on $\bar{\cal A}$. Then
at every vertex $v$ of $\gamma$, the following condition must
hold:
\begin{align}
n^i\sum_{v\in V(\gamma)}X^i_v\cdot \psi_\gamma=0.
\end{align}
where $X^i_v=\sum_{e\in E(\gamma),\,b(e)=v}X^i_e$. Under the
$n^i$-gauge fixing, the extrinsic scalar curvature $k$ reduces to
\begin{align}
k=\frac{n^i}{\sqrt{\det(q)}}\;\partial_aE^a_i.
\end{align}
Note that now $n^i$ is a non-dynamical constant which need not to
be quantized. For simplicity, we choose again adapted coordinates
$\{x^1,x^2,x^3\}$ with respect to $S$. Our aim is to quantize the
following quantity under the partial gauge fixing
\begin{align}
\label{E-1without}
 E_{Q,k}(S)&=-\frac{1}{\kappa}\int_S
             d^2x\frac{\sqrt{\det(\sigma)}(x)}{\sqrt{\det(q)}(x)}\;n^i(x)\partial_aE^a_i(y)\nonumber\\
          &=-\lim_{\epsilon\rightarrow 0}\frac{1}{\kappa}\int_S
             d^2x\frac{\sqrt{\det(\sigma)}(x)}{\sqrt{\det(q)}(x)}\;n^i(x)\int_\Sigma
             d^3y\frac{\chi^3_\epsilon(x,y)}{\epsilon^3}\;\partial_aE^a_i(y)\nonumber\\
          &=\lim_{\epsilon\rightarrow 0}\frac{1}{\kappa}\int_S
             d^2x\frac{\sqrt{\det(\sigma)}(x)}{V(x,\epsilon)}\;n^i(x)\int_\Sigma
             d^3y[\partial_a\chi^3_\epsilon(x,y)]E^a_i(y)\nonumber\\
          &=:\lim_{\epsilon\rightarrow 0}E^\epsilon_{Q,k}(S).
\end{align}
The first integration of Eq. (\ref{E-1without}) can be written as
\begin{align}
&\int_Sd^2x\,n^i(x)\frac{\sqrt{\det(\sigma)}(x)}{V(x,\epsilon)}=\int_Sd^2x\,n^i(x)
\sqrt{\frac{\tilde{n}_a(x)E^a_i(x)}{V(x,\epsilon)}\frac{\tilde{n}_b(x)E^b_i(x)}{V(x,\epsilon)}}\,,
\end{align}
where $\tilde{n}_a(x)=(dx^3)_a$. Using $g_{\epsilon'}(x,y)$
satisfying Eq. (\ref{geps}) as smearing function, we define
\begin{align}
\label{reg1}
\left[\frac{\tilde{n}_a(x)E^a_i(x)}{V(x,\epsilon)}\right]_{g_{\epsilon'}}&:=\int_Sd^2u\,g_{\epsilon'}(x,u)
\frac{\tilde{n}_a(u)E^a_i(u)}{V(u,\epsilon)}
=\int_S\frac{g_{\epsilon'}(x,u)}{V(u,\epsilon)}\frac{1}{2}\,\epsilon_{ijk}e^j(u)\wedge e^k(u)\nonumber\\
 &=\frac{8}{\kappa^2}\int_Sg_{\epsilon'}(x,u)\epsilon_{ijk}\{A^j(u),\sqrt{V(u,\epsilon)}\}
 \wedge\{A^k(u),\sqrt{V(u,\epsilon)}\},
\end{align}
where we have used the classical identity
\begin{align}
e_a^i(u)=\frac{2}{\kappa}\{A_a^i(u),V(u,\epsilon)\},
\end{align}
and absorbed $1/V(u,\epsilon)$ into the Poisson brackets. Thus we
have
\begin{align}
\label{reg2}
\int_Sd^2x\,n^i(x)\frac{\sqrt{\det(\sigma)}(x)}{V(x,\epsilon)}
=\lim_{\epsilon'\rightarrow0}\int_Sd^2x\,n^i(x)
\sqrt{\left[\frac{\tilde{n}_a(x)E^a_i(x)}{V(x,\epsilon)}\right]_{g_{\epsilon'}}
\left[\frac{\tilde{n}_b(x)E^b_i(x)}{V(x,\epsilon)}\right]_{g_{\epsilon'}}}\,\,.
\end{align}

We introduce a triangulation of the 2-d surface $S$
\cite{QSDIV,QSDVI}. Denote by $\Delta$ a solid triangle. Single out
one of the corners of the triangle and call it $v(\Delta)$. At
$v(\Delta)$ there are incident two edges $s_1(\Delta),s_2(\Delta)$
of $\partial \Delta$ which we equip with outgoing orientation, that
is, they start at $v(\Delta)$. Let us now write the integral over
$S$ as a sum of integrals over $\Delta$ where $\Delta$ are triangles
of some triangulation $T$ of $S$,
\begin{align}
\label{25}
{\left[\frac{\tilde{n}_a(x)E^a_i(x)}{V(x,\epsilon)}\right]_{g_{\epsilon'}}}&=\frac{4}{\kappa^2}
\sum_{\Delta\in T}g_{\epsilon'}(x,v(\Delta))\sum_{s_I(\Delta)\cap
s_J(\Delta)=v(\Delta)}
\epsilon(s_Is_J)\epsilon^{IJ}\epsilon_{ijk}\times\nonumber\\
&\;\;\mathrm{Tr}\left(\tau_jh_{s_I(\Delta)}\{h^{-1}_{s_I(\Delta)},\sqrt{V(v(\Delta),\epsilon)}\;\}\right)
\mathrm{Tr}\left(\tau_kh_{s_J(\Delta)}\{h^{-1}_{s_J(\Delta)},\sqrt{V(v(\Delta),\epsilon)}\;\}\right),
\end{align}
where $\epsilon(s_I
s_J):=\mathrm{sgn}(\det(\dot{s}_I\dot{s}_J)(v(\Delta)))$ takes the
values $+1, -1, 0$ if the tangents of the two segments $s_I,s_J$ at
$v$ (in that sequence) form a matrix of positive, negative or
vanishing determinant. For convenience, we introduce
\begin{align}
\label{e}
e_I^j(v(\Delta)):=\mathrm{Tr}\left(\tau_jh_{s_I(\Delta)}\{h^{-1}_{s_I(\Delta)},\sqrt{V(v(\Delta),\epsilon)}\;\}\right).
\end{align}
Then the quantity in the square root of the expression in
(\ref{reg2}) can be regularized as
\begin{align}
\label{regu1}
&{\left[\frac{\tilde{n}_a(x)E^a_i(x)}{V(x,\epsilon)}\right]_{g_{\epsilon'}}}
{\left[\frac{\tilde{n}_b(x)E^b_i(x)}{V(x,\epsilon)}\right]_{g_{\epsilon'}}}\nonumber\\
=&\frac{16}{\kappa^4} \sum_{\Delta,\Delta'\in T}
g_{\epsilon'}(x,v(\Delta))g_{\epsilon'}(x,v(\Delta'))\,\epsilon(s_I
s_J)\epsilon_{ijk}\epsilon^{IJ}
e_I^j(v(\Delta))e_J^k(v(\Delta))\nonumber\\
&\quad\times\,\epsilon(s_K
s_L)\epsilon_{imn}\epsilon^{KL}e^m_K(v(\Delta'))e_L^n(v(\Delta')).
\end{align}
We introduce a triangulation of the 2-d surface $S$ in adaption to
the graph $\gamma$ \cite{QSDIV,QSDVI} and only consider the terms in
(\ref{regu1}) which sum over triangles $\Delta$ whose
basepoint $v(\Delta)$ coincides with a vertex $v$ of the graph,
\begin{align}
\label{regu2} &\frac{16}{\kappa^4} \sum_{v,v'\in V(\gamma)\cap S}
g_{\epsilon'}(x,v)g_{\epsilon'}(x,v')\epsilon_{ijk}\epsilon_{imn}\sum_{s_I(\Delta)\cap
s_J(\Delta)=v} \frac{4}{E(v)}\,\epsilon(s_I s_J)\epsilon^{IJ}
e_I^j(v)e_J^k(v)\nonumber\\
&\quad\quad\quad\times\sum_{s_K(\Delta')\cap s_L(\Delta')=v'}
\frac{4}{E(v')}\,\epsilon(s_K s_L)\epsilon^{KL}e_K^m(v')e_L^n(v').
\end{align}
For sufficiently small $\epsilon'$,
$g_{\epsilon'}(x,v)g_{\epsilon'}(x,v')$ is zero unless $v=v'$. Then
the double sum over vertices reduces to a single one, Eq.
(\ref{regu2}) reduces to
\begin{align}
&\frac{16}{\kappa^4} \sum_{v\in V(\gamma)\cap S}
[g_{\epsilon'}(x,v)]^2\epsilon_{ijk}\epsilon_{imn}\frac{16}{E(v)^2}\nonumber\\
&\quad\times\sum_{s_I(\Delta)\cap s_J(\Delta)= s_K(\Delta)\cap
s_L(\Delta)=v} \epsilon(s_I s_J)\epsilon(s_K
s_L)\epsilon^{IJ}\epsilon^{KL}
e^j_I(v)e^k_J(v)e^m_K(v)e^n_L(v)\nonumber\\
=&\frac{32}{\kappa^4} \sum_{v\in V(\gamma)\cap S}
[g_{\epsilon'}(x,v)]^2\frac{16}{E(v)^2}\sum_{IJKL}\epsilon(s_I
s_J)\epsilon(s_K s_L)
\epsilon^{IJ}\epsilon^{KL}e_I^j(v)e_K^j(v)e_J^k(v)e_L^k(v).
\end{align}
where we have abbreviated $\sum_{s_I(\Delta)\cap s_J(\Delta)=
s_K(\Delta)\cap s_L(\Delta)=v}\rightarrow \sum_{IJKL}$. We further
introduce the manifestly gauge invariant quantities \cite{Tri}
\begin{align}
q_{IK}(v)=e_I^j(v)e_K^j(v).
\end{align}
The (\ref{regu2}) can be written as
\begin{align}
&\frac{32}{\kappa^4} \sum_{v\in V(\gamma)\cap S}
[g_{\epsilon'}(x,v)]^2\frac{16}{E(v)^2}\sum_{IJKL} \epsilon(s_I
s_J)\epsilon(s_K s_L)\epsilon^{IJ}\epsilon^{KL}q_{IK}(v)q_{JL}(v).
\end{align}
To write the regulated operator corresponding to (\ref{regu1}), we
replace Poisson brackets by commutators times $1/(i\hbar)$,
holonomies by multiplication and $V$ by $\hat{V}$. As we evaluate
the operator corresponding to Eq. (\ref{regu1}), we find out that
only those triangles $\Delta$ contribute whose basepoint $v(\Delta)$
coincides with a vertex $v$ of the graph due to the presence of the
volume operators. Hence we obtain the regulated operator
corresponding Eq. (\ref{regu1}) acting on a cylindrical function as
\begin{align}
\label{reguSq}
&\widehat{{\left[\frac{\tilde{n}_a(x)E^a_i(x)}{V(x,\epsilon)}\right]_{g_{\epsilon'}}}}
\widehat{{\left[\frac{\tilde{n}_b(x)E^b_i(x)}{V(x,\epsilon)}\right]_{g_{\epsilon'}}}}\cdot f_\gamma\nonumber\\
=&\frac{32}{\kappa^4}\cdot\frac{1}{(i\hbar)^4} \sum_{v\in
V(\gamma)\cap S} [g_{\epsilon'}(x,v)]^2\frac{16}{E(v)^2}\sum_{IJKL}
\epsilon(s_I s_J)\epsilon(s_K
s_L)\epsilon^{IJ}\epsilon^{KL}\hat{q}_{IK}(v)\hat{q}_{JL}(v)\cdot
f_\gamma\nonumber\\
=:&\sum_{v\in V(\gamma)\cap S}
[g_{\epsilon'}(x,v)]^2\,\hat{q}_v\cdot f_\gamma,
\end{align}
where
\begin{align}
\hat{q}_{IK}(v)=\hat{e}_I^j(v)\hat{e}_K^j(v)
\end{align}
with
\begin{align}
\hat{e}_I^i(v)=\mathrm{Tr}\left(\tau_ih_{s_I(\Delta)}
\left[h_{s_I(\Delta)}^{-1},\sqrt{\hat{V}_v}\;\right]\right)\Big|_{v\in
V(\gamma)}.
\end{align}
It is easy to see that the operator in (\ref{reguSq}) is gauge
invariant. Notice that the self-adjointness of $i\hat{e}_I^i(v)$
implies that $\hat{q}_v$ is a non-negative self-adjoint operator
and hence has a well defined square root. Since we have chosen
$\epsilon'$ to be sufficiently small, for any given point $x$ in
$S$, $g_{\epsilon'}(x,v)$ is non-zero for at most one vertex $v$.
We can therefore take the sum over $v$ outside the square root and
obtain
\begin{align}
&\left(\widehat{\left[\frac{\tilde{n}_a(x)E^a_i(x)}{V(x,\epsilon)}\right]_{g_{\epsilon'}}}
 \widehat{\left[\frac{\tilde{n}_b(x)E^b_i(x)}{V(x,\epsilon)}\right]_{g_{\epsilon'}}}\right)^{1/2}\cdot f_\gamma
=\sum_{v\in V(\gamma)\cap S}g_{\epsilon'}(x,v)\sqrt{\hat{q}_v}\cdot
f_\gamma.
\end{align}
Finally, we can remove the regulator, i.e., take the limit as
$\epsilon'$ tends to zero. Then the following equality holds in
the distributional sense.
\begin{align}
\label{distrioper}
\widehat{\left[\frac{\sqrt{\det(\sigma)}(x)}{V(x,\epsilon)}\right]}\cdot
f_\gamma=\sum_{v\in V(\gamma)\cap
S}\delta^2(x,v)[\hat{q}_v]^{1/2}\cdot
 f_\gamma.
\end{align}
The second integration of Eq. (\ref{E-1without}) can be similarly
quantized as
\begin{align}
\label{threedimoper} &\int_\Sigma
             d^3y[\partial_a\chi^3_\epsilon(x,y)]\hat{E}^a_i(y)\cdot f_\gamma\nonumber\\
             =&-\frac{i\hbar\kappa\beta}{2}\sum_{e\in E(\gamma)}\lim_{n\rightarrow
\infty}\sum_{k=1}^n\Big[\chi^3_\epsilon(x,e(t_k))-\chi^3_\epsilon(x,e(t_{k-1}))\Big]X^i_{e}(t_{k-1})\cdot
f_{\gamma}.
\end{align}
Using Eqs. (\ref{distrioper}) and (\ref{threedimoper}), we obtain
the regularized operator corresponding to $E_{Q,k}(S)$ as
\begin{align}
\hat{E}^{\epsilon,n}_{Q,k}(S)\cdot
f_\gamma&=-\frac{i\hbar\kappa\beta}{2}\int_S d^2xn^i(x)\sum_{v\in
V(\gamma)\cap S}\delta^2(x,v)[\hat{q}_v]^{1/2}\nonumber\\
  &\quad\quad\times\sum_{e\in E(\gamma)}\sum_{k=1}^n\Big[\chi^3_\epsilon(x,e(t_k))
  -\chi^3_\epsilon(x,e(t_{k-1}))\Big]X^i_{e}(t_{k-1})\cdot
    f_\gamma\nonumber\\
  &=-\frac{i\hbar\kappa\beta}{2}\sum_{v\in V(\gamma)\cap S}[\hat{q}_v]^{1/2}n^i(v)\nonumber\\
  &\quad\quad\times\sum_{e\in E(\gamma)}\sum_{k=1}^n\Big[\chi^3_\epsilon(v,e(t_k))
  -\chi^3_\epsilon(v,e(t_{k-1}))\Big]X^i_{e}(t_{k-1})\cdot
    f_\gamma.
\end{align}
Now we perform the limit $n\rightarrow\infty$ and
$\epsilon\rightarrow 0$ in reversed order. Keeping $n$ fixed, for
small enough $\epsilon$ only the term with $k =1$ in the sum
survives provided that $b(e)=v$. We then obtain the operator
\begin{align}
\label{hatE-2} \hat{E}_{Q,k}(S)\cdot f_\gamma
  &=\frac{i\hbar\kappa\beta}{2}\sum_{v\in V(\gamma)\cap S}[\hat{q}_v]^{1/2}n^i(v)\sum_{b(e)=v}X^i_{e}\cdot
    f_\gamma\nonumber\\
  &=-\hbar\kappa\beta\sum_{v\in V(\gamma)\cap S}[\hat{q}_v]^{1/2}n^i(v)\sum_{b(e)=v}Y^i_{e}\cdot
    f_\gamma,
\end{align}
where $Y_e^i\equiv-\frac{i}{2}X^i_e$ is the self-adjoint
right-invariant vector field. It is clear that the
$\hat{E}_{Q,k}(S)$ in (\ref{hatE-2}) is densely defined in ${\cal
H}_{\mathrm{kin}}$, and it vanishes the $n^i$-gauge-invariant
states.

\section{Proof of an equality}\label{proof}
We first give a proof for Eq. (\ref{Jident}) in \ref{section2}. By
choosing adapted coordinates $\{x^1,x^2,x^3\}$ with respect to $S$,
the normal-directional momentum of $S$ is given by
\begin{align}
\label{JQ1}
J_{Q,l}(S):=&\frac{1}{\kappa}\int_S d^2x\sqrt{\det(\sigma)}\;l\nonumber\\
   =&\frac{1}{\kappa}\int_S
   d^2x\sqrt{\tilde{n}_c\tilde{n}^c\det(q)}\,(K-n^an^bK_{ab})\nonumber\\
   \equiv&J_1(S)+J_2(S),
\end{align}
where
\begin{align}
l:=&\sigma^{ab}l_{ab}=\sigma^{cd}\nabla_cu_d=\nabla_au^a-n^cn^d\nabla_cu_d.\nonumber
\end{align}
The first term in Eq. (\ref{JQ1}) can be written as
\begin{align}
\label{J_1}
J_1(S)=\frac{1}{\kappa}\int_Sd^2x\sqrt{\tilde{n}_b\tilde{n}^b}\,E_i^aK_a^i,
\end{align}
and the second term as
\begin{align}
\label{J_2} J_2(S)
   =&-\frac{1}{\kappa}\int_Sd^2x\sqrt{\tilde{n}_c\tilde{n}^c}\,n^an_bK^i_aE^b_i.
\end{align}
Thus we obtain
\begin{align}
\label{J_l}
J_{Q,l}(S)=\frac{1}{\kappa}\Big(\int_Sd^2x\sqrt{\tilde{n}_b\tilde{n}^b}\,K_a^iE_i^a
-\int_Sd^2x\sqrt{\tilde{n}_c\tilde{n}^c}\,n^an_bK^i_aE^b_i\Big).
\end{align}
On the other hand,
\begin{align}
\label{A1} \{H^E(1),Ar(S)\}=&\kappa\int_\Sigma d^3y\frac{\delta
H^E(1)\delta
         Ar(S)}{\delta A^l_d(y)\delta E^d_l(y)}.
\end{align}
Note that the area of $S$ can be written as
\begin{align}
Ar(S)=&\int_Sd^2x\sqrt{\tilde{n}_a\tilde{n}_bE^a_iE^b_j\delta^{ij}}=\int_\Sigma
d^3x\sqrt{\tilde{n}_a\tilde{n}_bE^a_iE^b_j\delta^{ij}}\,\delta(x^3,0).
\end{align}
One gets
\begin{align}
\label{A2} \frac{\delta Ar(S)}{\delta
E^d_l(y)}=&\frac{\tilde{n}_bE^b_j\delta^{ij}\tilde{n}_a\delta^a_d\delta^l_i}
{\sqrt{\tilde{n}_e\tilde{n}_fE^e_mE^f_m\delta^{mn}}}\,\delta(y^3,0)
            =\frac{\tilde{n}_be^{bl}\tilde{n}_d}{\sqrt{\tilde{n}_a\tilde{n}^a}}\,\delta(y^3,0),\\
\frac{\delta H^E(1)}{\delta
                          A^l_d(y)}=&\,\frac{1}{\kappa}\left(\epsilon^{abd}\partial_ae_{bl}
                          +\epsilon^{abd}\epsilon_{ijl}A^i_ae^j_b\right).\label{A3}
\end{align}
Plugging Eqs. (\ref{A2}) and (\ref{A3}) into (\ref{A1}), we obtain
\begin{align}
\label{A4} \{H^E(1),Ar(S)\}=&\int_\Sigma
                          d^3y\frac{1}{\sqrt{\tilde{n}_e\tilde{n}^e}}(\epsilon^{abd}\partial_ae_{bl}
                          +\epsilon^{abd}\epsilon_{ijl}A^i_ae^j_b)\tilde{n}^ce_c^l\tilde{n}_d\delta(y^3,0).
\end{align}
The two terms in (\ref{A4}) can be reduced respectively to
\begin{align}
\epsilon^{abd}\partial_ae_{bl}\tilde{n}^ce_c^l\tilde{n}_d=&\epsilon^{[ab]d}(-{\Gamma^f}_{ab}e_{fl}
-\epsilon_{lmn}\Gamma^m_ae^n_b)\tilde{n}^ce_c^l\tilde{n}_d
     =-\Gamma^m_a\tilde{n}^c\tilde{n}_d\epsilon^{abd}\epsilon_{fbc}e^f_m\nonumber\\
                                         =&-\Gamma^m_aE^a_m\tilde{n}^d\tilde{n}_d
                                         +\Gamma^m_aE^d_m\tilde{n}_d\tilde{n}^a,\nonumber\\
\epsilon^{abc}\epsilon_{ijl}A^i_ae^j_b\tilde{n}^ce^l_c\tilde{n}_d
                  =&A^i_a\tilde{n}^c\tilde{n}_d\epsilon^{abc}\epsilon_{fbc}e^f_i
                  =A^m_aE^a_m\tilde{n}^d\tilde{n}_d-A^m_aE^d_m\tilde{n}_d\tilde{n}^a.
\end{align}
Hence one has
\begin{align}
\label{Poss} \big\{H^E(1),Ar(S)\big\}=&\int_\Sigma
d^3y\frac{1}{\sqrt{\tilde{n}_e\tilde{n}^e}}(K^m_aE^a_m\tilde{n}^d\tilde{n}_d
-K^m_aE^d_m\tilde{n}_d\tilde{n}^a)\delta(y^3,0)\nonumber\\
=&\int_S d^2yK^m_aE^a_m\sqrt{\tilde{n}^d\tilde{n}_d}-\int_S
d^2y\frac{K^m_aE^d_m\tilde{n}_d\tilde{n}^a}{\sqrt{\tilde{n}_e\tilde{n}^e}}\,.
\end{align}
The second integrand in (\ref{Poss}) reduces to
\begin{align}
\frac{K^m_aE^d_m\tilde{n}_d\tilde{n}^a}{\sqrt{\tilde{n}_e\tilde{n}^e}}
=&\sqrt{\tilde{n}_c\tilde{n}^c}\,n^an_bK^i_aE^b_i.\nonumber
\end{align}
We thus obtain
\begin{align}
\label{A5} \big\{H^E(1),Ar(S)\big\}=&\int_S
d^2y\sqrt{\tilde{n}^b\tilde{n}_b}K^i_aE^a_i-\int_S
d^2y\sqrt{\tilde{n}_c\tilde{n}^c}\,n^an_bK^i_aE^b_i.
\end{align}
Comparing (\ref{J_l}) to (\ref{A5}) we complete the proof.

\end{appendix}


\end{document}